\documentclass{article}
\usepackage{graphicx}
\def\encadre#1#2{%
\setbox100=\hbox{\kern#1{#2}\kern#1}
\dimen100=\ht100 \advance \dimen100 by #1
\dimen101=\dp100 \advance \dimen101 by #1
\setbox100=\hbox{\vrule height \dimen100 depth \dimen101\box100\vrule}
\setbox100=\vbox{\hrule\box100\hrule}
\advance \dimen100 by .4pt \ht100=\dimen100
\advance \dimen101 by .4pt \dp100=\dimen101
\box100
\relax
}

\newtheorem{thm}{{\sc Theorem}}
\newtheorem{cor}{{\sc Corollary}}
\newtheorem{fig}{{\sc Figure}}
\newtheorem{tab}{{\sc Table}}
\def\leurre{\noindent\leftskip0pt\small\baselineskip 10pt}
\def\cqfd{\hbox{\raise-2pt\hbox{\vbox to 12pt{\offinterlineskip
\hrule height 1pt depth 1pt width 12pt
\hbox to 12pt{\vrule height 4pt depth 4pt width 2pt
\hfill\vrule height 4pt depth 4pt width 2pt
}
\hrule height 1pt depth 1pt width 12pt
}}}}

\begin{document}
\def\ligne#1{\hbox to \hsize{#1}}
\def\PlacerEn#1 #2 #3 {\rlap{\kern#1\raise#2\hbox{#3}}}
\font\bfxiv=cmbx12 at 12pt
\font\bfxii=cmbx12
\font\pc=cmcsc10
\font\itix=cmti9
\font\rmix=cmr9 \font\mmix=cmmi9 \font\symix=cmsy9
\def\mathix{\textfont0=\rmix \textfont1=\mmix \textfont2=\symix}
\font\rmxiv=cmr14 \font\mmxiv=cmmi14 \font\symxiv=cmsy14
\def\mathxiv{\textfont0=\rmxiv \textfont1=\mmxiv \textfont2=\symxiv}
\font\rmvii=cmr7
\font\rmv=cmr5
\ligne{\hfill}
\pagenumbering{arabic}
\begin{center}\bf\Large
A note on groups of a family of hyperbolic tessellations
\end{center}
\vskip 10pt
\begin{center}\bf\large
Anthony Gasperin$^1$, Maurice Margenstern$^2$
\end{center}

\vskip 10pt
\begin{center}\small
$^1$
Laboratoire d'Informatique Th\'eorique, TCS,\\
Universit\'e de Gen\`eve.\\
{\it e-mail} : anthony.gasperin@unige.ch, anthony.gasperin@gmail.com
\vskip 5pt
$^2$
Laboratoire d'Informatique Th\'eorique et Appliqu\'ee, EA 3097,\\
Universit\'e de Lorraine, LITA EA3097,\\
Campus du Saulcy,\\
57045 Metz Cedex, France,\\
{\it e-mail} : maurice.margenstern@univ-lorraine.fr,margenstern@gmail.com
\end{center}

\vskip 10pt
\begin{abstract}
In this paper we study the word problem of groups corresponding to tessellations of the hyperbolic plane.
In particular using the Fibonacci technology developed by the second author we show that groups corresponding 
to the pentagrid or the heptagrid are not automatic.
\end{abstract}
\vskip 10pt



\section{Introduction}
The word problem for a finitely generated group is the algorithmic problem of deciding whether two words in the generators represent the same element.
Dehn in \cite{Dehn1911} suggested that it was an important area of study in its own right. In \cite{Dehn1912} he gave an algorithm that solves it 
for the fundamental group of closed orientable two-dimensional manifolds of genus greater or equal than two. Then the Dehn's algorithm has been extended and
applied to a wide range of group theoretic problems. In 1955 Novikov \cite{Novikov} showed that there exists a finitely presented group with an 
undecidable word problem.
A different proof is given by Boone in 1958. In cases where the problem is decidable it is interesting to study to which class the problem belongs. 
In particular it is knows that hyperbolic groups have word problem recognizable by a finite state automata. Groups with context-free word problem are 
precisely virtually free groups (\cite{Schu83},\cite{Schu85}) and their word problem is in fact deterministic context-free. 
Sapir, Birget and Rips obtained a characterization of groups with \textbf{NP} word problem: a finitely generated group has word problem in \textbf{NP} if 
and only if it embeds into a finitely presented groups with polynomial Dehn function.
In this paper we study the word problem of groups that corresponds to the pentagrid or the heptagrid. That is we consider the groups which have
a Cayley graph that represents the pentagrid or the heptagrid.
The study considers of the Fibonacci technology developed by the second author \cite{mmbook1},\cite{mmbook2}.
The technique allows one to find the location of a cell and its neighbours in 
the pentagrid. The Fibonacci technology has also some interesting algorithmic properties.
For example, in \cite{MMKGS13} an algorithmic description of contour words is emphasized and concluded using the Fibonacci technology.
We shall show that considering these algorithmic properties leads to a characterization of the groups we study.

\section{An abstract on hyperbolic geometry}
   
   In order to simplify the approach for the reader, we shall present a model
of the hyperbolic plane and simply refer to the literature for a more abstract,
purely axiomatic exposition. 

   As it is well known, hyperbolic geometry appeared
in the first half of the XIX$^{\rm th}$ century, in the last attempts to prove
the famous parallel axiom of Euclid's {\it Elements} from the remaining ones.
Hyperbolic geometry was yielded as a consequence of the repeated failure of
such attempts. Lobachevsky and, independently, Bolyai, 
discovered a new geometry by assuming that in the plane, from a point out of a 
given line, there are at least two lines which are parallel to the given line. 
Later, during the XIX$^{\rm th}$ century, models were discovered that gave 
implementations of the new axioms. The constructions of the models, all 
belonging to Euclidean geometry, proved by themselves that the new axioms bring
no contradiction to the other ones. Hyperbolic geometry is not less sound than 
Euclidean geometry is. It is also no more sound, in so far as much later, 
models of the Euclidean plane were discovered in the hyperbolic plane.

   Among the models of hyperbolic geometry, Poincar\'e's models met with great success because in 
these models, hyperbolic angles between lines coincide with the Euclidean 
angles of their supports. In this paper, we take Poincar\'e's disc as a 
model of the hyperbolic plane.

\subsection{Lines of the hyperbolic plane and angles}

   In Poincar\'e's disc model, the hyperbolic plane is the set of points
lying inside a fixed open disc of the Euclidean plane. 
The points which are on the border of the disc do not belong to the hyperbolic plane. However
they play an important role in the model and they are called {\bf points at infinity}. We shall
also call their set the {\bf border circle}.

   The lines of the hyperbolic plane in Poincar\'e's disc model are the trace in the disc of
or circles, orthogonal to the border circle. We say that
such a circle {\it supports} the hyperbolic line, $h$-line
for short, and sometimes simply {\it line} when there is no ambiguity.
\vskip 8pt
   Poincar\'e's unit disc model of the hyperbolic plane makes an intensive
use of some properties of the Euclidean geometry of circles, see \cite{MMjucs}
for an elementary presentation of the properties which are needed for our paper.
\vskip 5pt
\vskip 5pt
It is easy to see that an $h$-line defines two
points at infinity by the intersection of its Euclidean support with the
unit circle. They are called points at infinity of the $h$-line. The following
easily proved properties will often be used: 
any $h$-line has exactly two points at infinity; 
two points at infinity define a unique $h$-line passing through them; 
a point at infinity and a point in the hyperbolic plane 
uniquely define a $h$-line.
   
\vskip 4pt

   The angles between $h$-lines are defined as the Euclidean angle between 
the tangents to the arcs which are taken as the support of the corresponding 
$h$-lines. This is one reason for choosing that model: hyperbolic angles 
between $h$-lines are, in a natural way, the Euclidean angle between the 
corresponding supports. In particular, orthogonal circles support
perpendicular $h$-lines.

   In the hyperbolic plane, given a line, say $\ell$,  and a point~$A$ not 
lying on $\ell$, there are infinitely many lines passing through~$A$ which
do not intersect $\ell$. In the Euclidean plane, two lines are parallel if 
and only if they do not intersect. If the points at infinity are added to 
the Euclidean plane, parallel lines are characterized as the lines passing 
through the same point at infinity. Hence, as for lines, to have a 
common point at infinity and not to intersect is the same property in the 
Euclidean plane.

\vskip 10pt
\setbox110=\hbox{\includegraphics[scale=1.25]{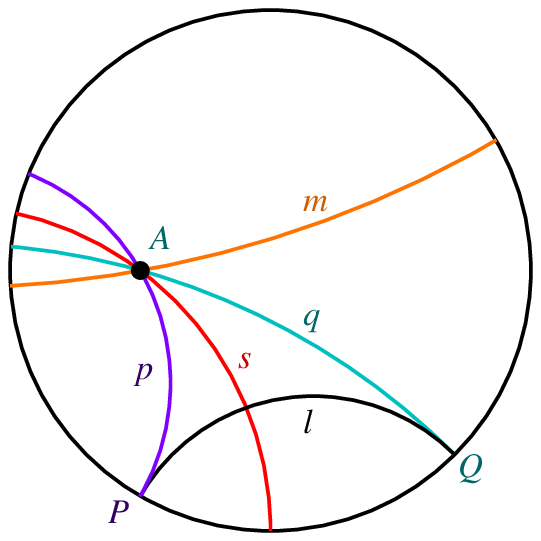}}
\vtop{
\ligne{\hskip 20pt
\box110
\hfill}
\vskip 0pt
\begin{fig}\label{poincare}
\leurre
The $h$-lines $p$ and $q$ are parallel to the $h$-line~$\ell$, with points at infinity~$P$
and~$Q$. The $h$-line~$s$ cuts~$\ell$ and the $h$-line~$m$ is non-secant with~$\ell$.
\end{fig}
\vskip 10pt
}
 
This is not the case in the hyperbolic plane, where two lines 
may not intersect and have no common point at infinity. We shall distinguish 
these two cases by calling {\bf parallel}, \hbox{$h$-lines} that share a common 
point at infinity, and {\bf non secant}, $h$-lines which have no common point 
at all neither in the hyperbolic plane nor at infinity. So, considering the 
situation illustrated by Figure~\ref{poincare}, there are exactly two $h$-lines 
parallel to a given $h$-line~$\ell$ which pass through a point~$A$ not lying on~$\ell$ 
and infinitely many ones which pass through~$A$ but are non-secant with~$\ell$. 
This is easily checked in Poincar\'e's disc model, see Figure~\ref{poincare}. Some authors call 
{\it hyperparallel} or {\it ultraparallel} lines
which we call {\it non-secant}.

   Another aspect of the parallel axiom deals with the sum of interior 
angles at the vertices of a polygon. In the Euclidean plane, the sum
of angles of any triangle is exactly $\pi$. In the hyperbolic plane, this 
is no more true: the sum of the angles of a triangle is {\it always 
less} than $\pi$. The difference from $\pi$ is, by definition, the 
{\bf area} of the triangle in the hyperbolic plane. Indeed, one can see
that the difference of the sum of the angles of a triangle from $\pi$ has the 
additive property of a measure on the set of all triangles. As a consequence, 
there is no rectangle in the hyperbolic plane. Consequently two non-secant 
lines, say $\ell$ and $m$, have, at most, one common perpendicular. It can be 
proved that this is the case: two non-secant lines of the hyperbolic 
plane have exactly one common perpendicular. 
\def\Mesh{[]}  

   It can be added that parallel $h$-lines have no common perpendicular.
   Consider the following problem of Euclidean geometry:
\vskip 7pt
\ligne{
$(P)$\hfill $\vcenter{\vbox{\parindent0pt\leftskip 0pt
                            \hsize=310pt
                            Let $\alpha,\beta,\gamma$ be positive real
                            numbers such that 
                            $\alpha$+$\beta$+$\gamma$ = $\pi$ and let
                            be given two lines $\ell$, $m$ intersecting
                            in $A$ with angle $\alpha$. How many 
                            triangles $ABC$ can be constructed with 
                            $B\in\ell$, $C\in m$ and $BC$ making angle
                            $\beta$ in~$B$ with $\ell$?
                            }%
                       }$
\hfill}
\vskip 5pt
   The answer is clearly: infinitely many. That property of 
the Euclidean plane defines the notion of {\it similarity}.
    
\vskip5pt
   Another consequence of the non-validity of Euclid's axiom on 
parallels in the hyperbolic plane is that there is no notion of
similarity in that plane: {\it if $\alpha,\beta,\gamma$ are 
positive real numbers such that $\alpha$+$\beta$+$\gamma < \pi$, $\ell$ 
and $m$ are $h$-lines intersecting in $A$ with angle $\alpha$, there are 
exactly two triangles $ABC$ such that $B\in\ell$, $C\in m$ and $BC$ makes 
angle $\beta$ in~$B$ with $\ell$ and angle $\gamma$ in~$C$ with $m$}. 
Each of those triangles is determined by the side of $\ell$ with respect
to $A$ in which $B$ is placed.
 
\subsection{Reflections in a $h$-line}

   Any $h$-line, say $\ell$, defines a {\bf reflection} in that line denoted 
by $\rho_{\ell}$. Let~$\Omega$ be the center of the Euclidean support of 
$\ell$, $R$ its radius. Two points~$M$ and $M'$ are {\bf symmetric} with 
respect to $\ell$ if and only if $\Omega$, $M$ and $M'$ belong to the same 
Euclidean line and if $\Omega M.\Omega M' = R^2$. Moreover, $M$ and $M'$ do 
not lie in the same connected component of the complement of~$\ell$ in the unit
disc. We also say that $M'$ is obtained from $M$ by the reflection in~$\ell$. 
It is clear that $M$ is obtained from $M'$ by the same reflection.

   All the transformations of the hyperbolic plane that we shall later consider
are reflections or constructed as products of reflections. 

   By definition, an {\bf isometry} of the hyperbolic plane is a finite
product of reflections. Two segments $AB$ and $CD$ are called 
{\it equal} if and only if there is an isometry transforming $AB$ 
into~$CD$.

   It is proved that finite products of reflections can be characterized
as either a single reflection or the product of two reflections or the
product of three reflections. In our sequel, we will mainly be 
interested by single reflections or products of two reflections.
We shall briefly mention products of three reflections in section~\ref{automaticity}.
The set which contains the identity and the product of two reflections 
constitutes a group which is called the {\bf group of motions}.

   At this point, we can compare {\it reflections} in a line in the 
hyperbolic plane with symmetries with respect to a line in the Euclidean
plane. Indeed, these respective transformations share many properties
on the objects on which they respectively operate. However, there is a 
very deep difference between the isometries of the Euclidean plane and 
those of the hyperbolic plane: while in the first case, the group of 
motions possesses non trivial normal subgroups, in the second case, 
this is no more the case: the group is simple.

   The product of two reflections with respect to lines~$\ell$ and~$m$
is a way to focus on that difference. In the Euclidean case, according
to whether~$\ell$ and~$m$ do intersect or are parallel, the product of
the two corresponding symmetries is a rotation around the point of 
intersection of~$\ell$ and~$m$, or a shift in the direction perpendicular
to both~$\ell$ and~$m$. In the hyperbolic case, if $h$-lines $\ell$ 
and~$m$ intersect at a point $A$, the product of the corresponding 
reflections is again called a rotation around $A$ as far as the obtained
transformation can be considered as what is intuitively called a 
rotation. But, if $\ell$ and~$m$ do not intersect, there are two 
cases: either $\ell$ and $m$ intersect at infinity, or they do not 
intersect at all. This gives rise to different cases of shifts. The 
first one, called {\bf ideal rotation}, is a kind of 
degenerated rotation, as in the Euclidean case, and the second one is 
called {\bf hyperbolic shift} or simply {\bf shift} along the common 
perpendicular to~$\ell$ and~$m$. Such a shift can be characterized by 
the image~$P'$ of any point~$P$ on the common perpendicular, say~$n$. We shall 
speak of the shift along~$n$ transforming~$P$ into~$P'$. When~$n$, $P$ and~$P'$ are
clear from the context, they are omitted and we simply speak about the shift.

   It can be proved that for any couple of two $h$-lines $\ell$ and~$m$,
there is an $h$-line~$n$ such that~$\ell$ and~$m$ are exchanged in the
reflection in~$n$. In the case when $\ell$ and~$m$ are non-secant,
$n$ is the perpendicular bisector of the segment that joins the 
intersections of $\ell$ and~$m$ with their common perpendicular. 

\def\reunion{\mathop{\cup}}
\section{The group-theoretic approach}

    It is well known that the notion of reflection in the Euclidean geometry
is an important case which illustrates the power of the theory of groups which 
operate on a space.

   An important notion which applies this general idea to many geometrical
settings is the notion of Cayley graph.

   Consider a directed graph $(V,A)$, where $V$ is the set of vertices and 
$A$ is the set of arcs and assume that the degree of the graph is constant 
and finite. We say that $(V,A)$ is the {\bf Cayley graph} of some group $G$ if 
and only if $G$ is finitely generated, and there is a labelling of the arcs by 
generators of $G$ or their inverses such that:
\vskip 5pt
{\leftskip 40pt\parindent0pt
- for any arc $\alpha$ from $v$ to $w$, $\gamma$ labels $\alpha$ if and only
if $\gamma(v) = w$;

- for any vertex~$v$, the labels which are set on the arcs starting from $v$
or arriving to~$v$ generate~$G$; they constitutes the 
{\bf labels around $v$};

- the set of labels around $v$ is the same for all the vertices of $V$.
\par}
\vskip 5pt
   A {\bf Cayley graph} is a directed graph $(V,A)$ for which there is a 
finitely generated group $G$ such that $(V,A)$ is the Cayley graph of 
$(V,A)$.

   The three regular tilings of the Euclidean plane are examples of
Caley graphs of groups. More important, they are Cayley graphs of groups
which are closely related with the motions which leave the tiling
invariant.

   A classical example is the example of the hexagonal tiling of the
Euclidean plane. Fix~$A$ a vertex and call $a$, $b$, $c$ the three shifts which 
transform~$A$ into the other end of the arc starting from~$A$ or arriving to~$A$. 
It is not difficult to see that we can identify any vertex by a word on 
the alphabet $\{a,b,c,a^{-1},b^{-1},c^{-1}\}$, and we have a simplification
rule given by $ab^{-1}ca^{-1}b = 1$, where 1 denotes the identity 
transformation.

   It is plain that going from any vertex~$v$ to another one~$w$ defines a 
word whose letters label the {\bf path} from~$v$ to~$w$. It is no less plain
that there are infinitely many words that can be associated in that way for a
given couple of points and even a big number of them that have the shortest
length. And so, the problem to decide whether or not two words define the
same point starting from a given point and following the paths that the words
label is an important one. This problem is well know as the word problem for 
groups. It is known to be undecidable in general for finitely generated groups 
with a finite number of identifying relations, see \cite{Novikov}.
Fortunately, in our case, the group is decidable. Indeed, it is the
case for most of the Cayley graphs used in geometry. Moreover, in many cases, 
there is a finite automaton which recognizes whether a word defines a cycle or not.
From that, it is easy to devise an automaton which recognizes whether two
words define or not the same end point if starting from the same point.
Together with another property of groups that goes far beyond the scope
of this paper, such groups are called {\bf automatic}. And so, the tilings
of the Euclidean plane are Cayley graphs of reflection groups of the
Euclidean plane which are automatic. Important works in this line has been 
done by Epstein, Thurston et all, see \cite{Epstein92}.

   Unfortunately, this situation does not fully extends to the hyperbolic
plane.

   The first reason is that, in the hyperbolic plane, there are infinitely
many regular tilings. This is a consequence of a well known theorem due to 
Poincar\'e. But before stating the theorem, we need to remind a few
definitions.

  Tessellations in the plane $-$ the definition is independent of the geometry
that we consider $-$ consist in the following operations. First, take a convex 
polygon~$P$. Let ${\cal S}(P)$ be the set of the lines that support its sides. 
If $\cal E$ is a set of polygons, one extends $\cal S$ to $\cal E$ by setting 
that ${\cal S}({\cal E}) = 
\displaystyle\reunion\limits_{P\in{\cal E}}{\cal S}(P)$.
Given $\cal K$ a set of lines and $\cal E$ a set of polygons, we define that 
$\rho_{\cal K}({\cal E}) =
\displaystyle\reunion\limits_{k\in{\cal K},Q\in{\cal E}}\rho_k(Q)$. 
Setting
${\cal T}_0 = \{P\}$, we inductively define ${\cal T}_{k+1}$ by
${\cal T}_{k+1} = \rho_{{\cal S}({\cal T}_k)}({\cal T}_k)$. Finally, we define 
${\cal T}^* = \displaystyle\reunion\limits_{k=0}^{\infty}{\cal T}_k$ to
be the {\it tessellation} generated by $P$. We say that the tessellation
is a {\it tiling} if and only if the following conditions hold:
\vskip 7pt
{\leftskip 20pt
- any point of the plane belongs to at least one polygon in
${\cal T}^*$;

- the interiors of the elements of ${\cal T}^*$ are pairwise disjoint.
\par}
\vskip 3pt
   In that definition, lines are defined according to the considered geometry.
It may have a consequence on the existence of a tessellation, depending on
which polygon is taken in the first step of the construction. As an example,
starting from a regular figure, there are three possible tessellations
giving rise to a tiling of the Euclidean plane, up to similarities, the three regular tilings which 
we already know as examples of Cayley graphs.
\vskip 5pt
    We can now state the theorem:
\vskip 5pt
\begin{thm}\label{thmpoincare}
{\bf Poincar\'e's Theorem}, {\rm (\cite{Poin})} $-$ Any triangle with 
angles $\displaystyle{\pi/\ell}$,
$\displaystyle{\pi/m}$ and $\displaystyle{\pi/n}$ such that 
\vskip 5pt
\ligne{\hfill
$\displaystyle{{1\over\ell}+{1\over m}+{1\over n}} < 1$ \hfill}

\noindent
generates a unique tiling by tessellation.
\end{thm}

   As an immediate corollary of the theorem, tilings based on any regular
polygon with $p$ sides and interior angle ${2\pi}\over q$ do exist, provided
that $\displaystyle{{1\over p} + {1\over q} < {1\over 2}}$. Such a polygon and
the corresponding tiling are denoted by $\{p,q\}$. Now, it is known, see 
\cite{Cox} that the Cayley graph of the graph generated by the reflections 
in the sides of the polygon $\{p,2q\}$ is the tiling $\{2q,p\}$, which is
the dual graph of the tiling $\{p,2q\}$. But the theorem says nothing for
tilings $\{s,r\}$ with an odd $s$. In the example of the pentagrid that we 
shall later more closely study, {\it i.e.} the tiling $\{5,4\}$, it can be
shown that this tiling is the Cayley graph of some abstract group but it is
not the Cayley graph of any group of isometries of the hyperbolic plane.  

\section{Locating the tiles in a tessellation of the hyperbolic plane}

\subsection{The pentagrid}
\vskip 5pt
 As we already noticed in the previous section, Poincar\'e's theorem 
immediately shows that a tiling is generated by tessellation if we take
the triangle with the following angles : 
$\displaystyle{{\pi\over5},{\pi\over4},{\pi\over2}}$. It is easy to see
that ten of those triangles share the same vertex corresponding to the angle
$\displaystyle{\pi\over5}$ and that such a grouping defines a regular pentagon
with right angles. This tiling is classically denoted by $\{5,4\}$ and we shall
from now on call it the {\bf pentagrid}, a representation of which in the 
south-western quarter of the hyperbolic plane is shown below in Figure~\ref{camembert}. 

It should be noticed that the pentagrid is the simplest regular grid of the 
hyperbolic plane. The triangular equilateral grid and the square grid of the 
Euclidean plane cannot be constructed here as they violate the law about the 
sum of angles in a triangle which is always less than $\pi$ in the hyperbolic 
plane.

   Poincar\'e's theorem was first proved by Henri Poincar\'e, \cite{Poin},
and other proofs were given later, for example in \cite{Cara}. In 
\cite{MMKM98}, another proof is provided for the existence
of the pentagrid which gives rise to a feasible algorithm in order to locate 
cells. The next section is devoted to a short presentation of that new proof.

\vtop{
\vskip 10pt
\ligne{\hskip20pt
\includegraphics[scale=1.3]{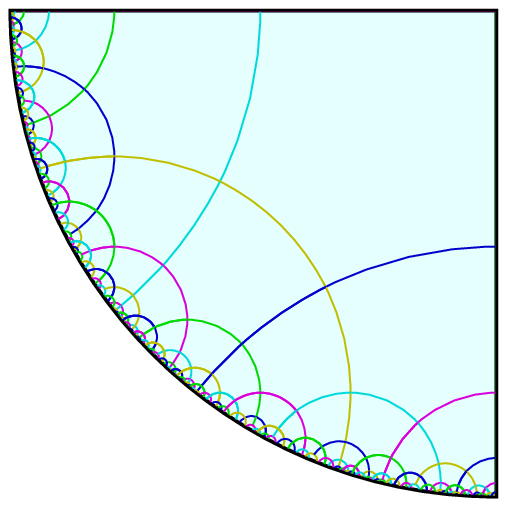}
\hfill}
\vskip 8pt
\begin{fig}\label{camembert}
\leurre
The pentagrid in the south-western quarter
\end{fig}
\vskip10pt
}

   It is based on a recursive splitting process that is illustrated by 
Figure~\ref{split54}, below.

   The  proof given in \cite{MMKM98}\ constructs a bijection between the tiling 
of the south-western quarter of the hyperbolic plane, say $\cal Q$, and a 
special infinite tree: the {\it Fibonacci tree}. Notice that $\cal Q$ is
isometric to any quarter of the hyperbolic plane.

\vskip 5pt
   Here, we remind sketchily the construction of that bijection.

\vskip 5pt
   Let $P_0$ be the regular rectangular pentagon contained in $\cal Q$ that 
has one vertex on the center of the unit disc and two sides supported by the 
sides of $\cal Q$. Say that $P_0$ is the {\it leading} pentagon of $\cal Q$.

\def\E#1{{\bf #1}}
 Number the sides of $P_0$ clockwise by \E1, \E2, \E3, \E4\ and \E5\ as 
indicated below, on Figure~\ref{split54}. As \E1\ is perpendicular to \E2\ and \E5\ and
as \E4\ is perpendicular to \E3\ and \E5, \E2\ and \E3\ do not intersect \E5.
The complement of $P_0$ in $\cal Q$ can be split into three regions as follows.
Line \E2\ splits $\cal Q$ into two components, say $R_1$ and $R'_1$ with $R'_1$
containing $P_0$. Line \E3\ splits $R'_1$ into $R_2$ and $R'_2$ with $R'_2$
containing $P_0$. Line \E4\ splits $R'_2$ into $P_0$ and $R_3$.
This defines the initial part of a tree: $P_0$ is associated to the root of 
the tree, and let us consider that the root has three sons, 
ordered from left to right and respectively associated to $R_3$, 
$R_2$ and~$R_1$. We can denote it as indicated by Figure~\ref{split54}.
We shall say that the root is a 3-node because it has three sons. 
\vskip 10pt
\setbox110=\hbox{\includegraphics[scale=1.2]{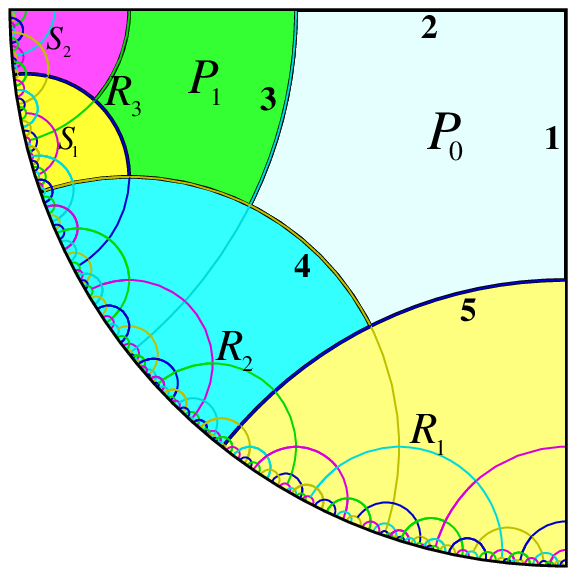}}
%
\vtop{
\ligne{\hfill 
\box110\hfill}
\vskip0pt
\begin{fig}\label{split54}
\leurre
Splitting the quarter into four parts
\vskip 2pt
First step: regions $P_0$, $R_1$, $R_2$ and $R_3$, where region $R_3$ is 
constituted of regions $P_1$, $S_1$ and $S_2$;
\vskip 1pt
Second step: regions $R_1$ and $R_2$ are split as the quarter (not represented)
while region $R_3$ is split into three parts: $P_1$, $S_1$ and $S_2$ as 
indicated in the figure.
\end{fig}
\vskip 9pt
}

Regions $R_1$ and $R_2$ are isometric images of $\cal Q$ by simple 
displacements: $R_1$ is obtained from $\cal Q$ by the displacement along 
\E1\ that transforms \E5\ into~\E2. Similarly for $R_2$ with the displacement 
along \E4\ that transforms \E5\ into~\E3. The same splitting into four parts 
can be repeated for these regions. Their leading pentagons are also 3-nodes.

   Now, let us see the status of region~$R_3$. It is plain that $R_3$
{\it is not} isometric to $\cal Q$, it is a {\it half-strip}: by analogy with
Euclidean geometry, we call here {\it strip} a region that is delimited by two
non-secant lines. Both {\it half-strips} are delimited by the common 
perpendicular of the lines. Let $P_1$ be the reflection of $P_0$ in \E4\ with 
sides which are now numbered anti-clockwise, so that the same number is given 
to the edges supported by the same $h$-line. In order to avoid possible 
confusion, we put the name of the considered pentagon as an index, if needed. 
Say that $P_1$ is the leading pentagon of~$R_3$. 
Notice that $R_3\cup P_0$ is transformed into a region $\cal S$ 
by the 
displacement 
along \E5\ that transforms \E1$_{P_0}$ into \E4$_{P_0}$, say
$\Delta$, see Figure~\ref{split54}. 
Define $S_1$ and $S_2$ as the respective images of $R_2$ and $R_3$ by $\Delta$. 
Then notice that ${\cal S} = S_2\cup P_1$. Say that $S_1$ and $S_2$ are the 
sons of $R_3$ and associate also these nodes to their leading pentagon. We say 
that the node associated to $R_3$ is a 2-node.
\vskip5pt
\vskip 4pt
   One can clearly see how we may proceed now. Define the following two 
rules:

{\leftskip 20pt \parindent 0pt
- a 3-node has three sons: to left, a 2-node and, in 
the middle and to right, in both cases, 3-nodes;

- a 2-node has 2 sons: to left a 2-node, to right a 3-node.
\par}
\vskip3pt
\noindent
Those two rules, combined with the axiom which tells that the root is a 
3-node, uniquely define a tree which we call the {\it standard Fibonacci} tree, 
see Figure~\ref{fibotree}, below. 

   The properties of the standard Fibonacci tree are indicated in 
\cite{MMKM98}, \cite{MMKM99}{} and \cite{MMKMTCS}, and they are thoroughly 
proved in \cite{MM99}. Important works were performed after these pioneer work,
they are presented, most of them in full details, in \cite{mmbook1,mmbook2,mmbook3}.
Here, our attention is focused on on the {\it location} of the elements of the pentagrid
which will allow us to give the tools for Section~\ref{automaticity}.

\subsection{The Fibonacci technology}

   In principle, the technique that is used in \cite{MMKM98,MMKM99} and 
\cite{MMKMTCS} allows us to find the location of a cell and its neighbours in 
the pentagrid. This is the reason why the quoted papers assume that the
Fibonacci tree is implemented in the hardware of the cellular automaton. A way
to implement the tree consists in assuming that the path from the root to each 
node is known: it may be stored as a sequence of the sides (numbered from 1 to 
5) in which reflections are performed starting from the pentagon of the 
root until the right node is reached. In order to locate the neighbours of the 
cells in that setting, it can be assumed that for each node, the path to the 
next node on the same level is also given. Otherwise, it would be possible to 
compute it, but at the price of a complete induction.
   
   In \cite{MM99,MMjucs}, a new and more efficient way is defined
to locate the cells which lie in the quarter, by numbering the nodes of the 
tree with the help of the positive numbers. We attach 1 to the root and then, 
the following numbers to its sons, going on on each level from left to right 
and one level after another one, see Figure~\ref{fibotree}, below. 
\vskip 4pt
   That numbering is fixed once and for all in the paper. We fix also a 
representation of the numbers by means of the Fibonacci sequence, 
$\{f_i\}_{i\in I\!\!N}$, which is defined by the induction relation
$f_{n+2} = f_{n+1}+f_n$, and $f_0=f_1=1$.

\vtop{
\vskip 10pt
\ligne{
\includegraphics[scale=1.2]{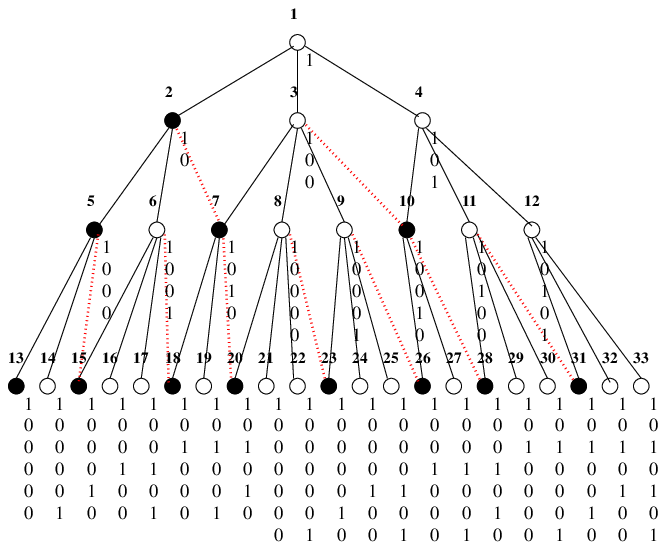}
\hfill}
\vskip 8pt
\begin{fig}\label{fibotree}
\leurre
The standard Fibonacci tree: 
\vskip0pt above a node: its number; below: its standard representation.
\vskip 0pt Notice that the first node of a level has a Fibonacci number with
odd index as its number. The number of nodes on a level is also a Fibonacci
number. This property is the reason why the tree is called a {\it Fibonacci} 
tree.
\end{fig}
\vskip 10pt
}

   It is known that every positive number $n$ is a sum of distinct Fibonacci 
numbers: $n=\displaystyle{\sum\limits_{i=1}^k\ \alpha_i.f_i}$ with 
$\alpha_i\in\{0,1\}$.  Such a representation defines a word
 $\alpha_k\ldots\alpha_1$ which is called a {\it Fibonacci representation} 
of~$n$.

It is known that such a representation is not unique, but it can be made
unique by adding a condition. Namely, we can assume that in the representation,
there is no occurrence of the pattern 11: if $\alpha_i = 1$ in
the above word, then $i = k$ or $\alpha_{i+1}=0$. Following \cite{MM99,MMjucs},
we shall say that this new representation is the {\it standard} one. In 
\cite{MM99,MMjucs}, we give a proof of these well-known features.

   From the standard representation, which can be computed in linear time
from the number itself, see \cite{MMjucs}, it is possible to find the 
information that we need to locate the considered node in the tree: we can 
find its {\it status}, {\it i.e.} whether it is a 2-node or a 3-node; the 
number of its father; the path in the tree that leads from the root to that 
node; the numbers attached to its neighbours. All this information can be
computed in linear time.
This is done in great detail in \cite{MM99}\ for the considered tree. 
\vskip 5pt
   In \cite{MMjucs}, we proved that there are many other Fibonacci trees.
There is indeed a continuous family of them.
 
  In order to see that, consider again Figure~\ref{split54}. Indeed, that figure
contains all the information that is needed in order to state the rules
that lead to the tree represented in Figure~\ref{fibotree}.

\vskip 5pt
   Now, we can split the quarter in another way, as shown by 
Figure~\ref{splitcentre}, below.

\vtop{
\vskip 5pt
\setbox110=\hbox{\includegraphics[scale=0.5]{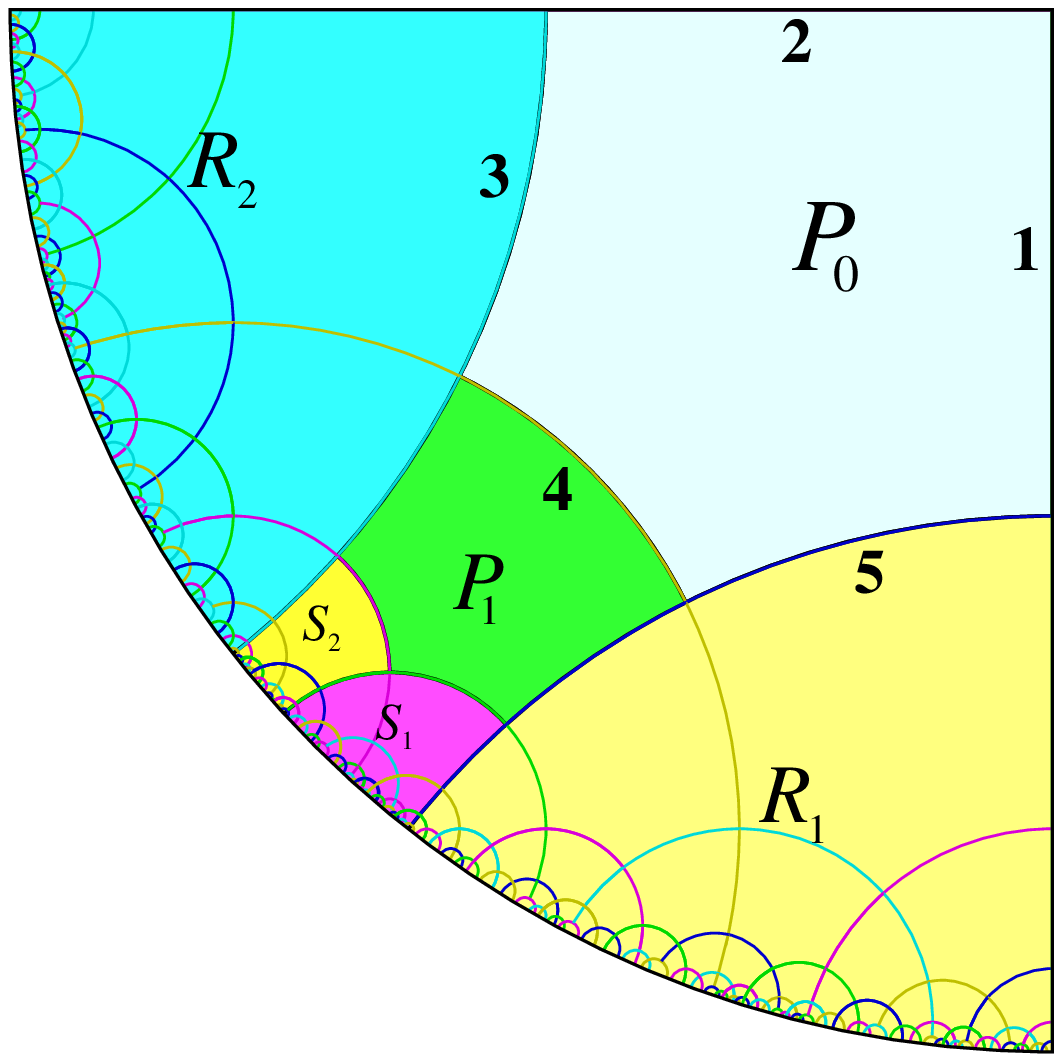}}
\ligne{\hfill
\box110\hfill}
\vskip 0pt
\vspace{-10pt}
\begin{fig}\label{splitcentre}
\leurre
Splitting the quarter into four parts in another way
\vskip0pt
Region $R_3$ consists of $P_1$, $S_1$ and $S_2$.
\end{fig}
\vskip 8pt
}

   This defines a new splitting which differs from the one defined in 
\cite{MMKM98,MMKM99,MMKMTCS} and \cite{MM99}, only on the way with which
the regions that are isometric to a quarter are chosen.

   At this point, we can notice that we can apply the arguments given in
\cite{MMKM98,MMKMTCS,MM99} in order to prove the bijection between the new
tree and the tiling of the quarter. Indeed, when we consider the diameter of 
a region that tends to zero as the index of the step of splitting tends to 
infinity, the estimates that we then established are still in force here. 
\vskip 5pt
   Let us now focus on the trees that are obtained. The standard Fibonacci 
tree defined in the first papers can be rewritten as indicated in Figure~\ref{fibotree} 
where the numbers of the nodes are also displayed with their standard 
representation.
\vskip 5pt

   The new splitting that we define with the help of Figure~\ref{splitcentre} gives rise to a 
new kind of Fibonacci tree, where the rules for the nodes are different for 
the 3-nodes. In the case of the standard Fibonacci tree, the rules can also 
be expressed as follows: {\bf 2} $\rightarrow$ {\bf 2}$\,${\bf 3} and 
{\bf 3} $\rightarrow$ {\bf 2}$\,${\bf 3}$\,${\bf 3}. In 
the case of this new tree, let us call it {\it central} Fibonacci tree, the 
rules are: {\bf 2} $\rightarrow$ {\bf 2}$\,${\bf 3} and 
{\bf 3} $\rightarrow$ {\bf 3}$\,${\bf 2}$\,${\bf 3}. 
\vskip 5pt
   As already indicated, the numbering of the nodes in the tree is fixed and so,
the standard representation fixes the chosen Fibonacci representation. However,
the algorithms which gives the status of a node, the number of the father, the 
path from the root to the considered node and the numbers of its neighbours 
will be different, see \cite{MM2000,MMjucs} for more details.
\vskip 5pt
There are infinitely many general Fibonacci trees. They can be all constructed 
by a random algorithm using a dice\footnote{We use a {\itix cubic}, hence 
Euclidean, dice in a three-dimensional Euclidean space.} as follows:
\vskip 7pt
{\leftskip 20pt\parindent 0pt
- construct the root as a 3 node, which is at level 0;
\vskip 0pt
- iteratively construct levels one after another:

\leftskip 30pt
for each node of the current level:

\leftskip 40pt
throw the dice and let $r$ be the result

for a 2-node apply rule {\bf 2} $\rightarrow$ {\bf 2}$\,${\bf 3} iff $r <4$, 
otherwise {\bf 2} $\rightarrow$ {\bf 3}$\,${\bf 2}

\leftskip 60pt\parindent-20pt
for a 3-node apply rule {\bf 3} $\rightarrow$ {\bf 2}$\,${\bf 3}$\,${\bf 2} 
iff $r <3$, else {\bf 3} $\rightarrow$ {\bf 3}$\,${\bf 2}$\,${\bf 3} if 
$r < 5$, otherwise {\bf 3} $\rightarrow$ {\bf 3}$\,${\bf 3}$\,${\bf 2}.
\par}
\vskip 5pt
   As we have only permutations in the position of the 2-node among the sons
of a node, this does not change the number of nodes which occur and, by 
induction on the level of the considered tree, it is easy to see that
the number of nodes in a considered level is always the same for any
general Fibonacci tree. Consequently, the numbering is always the same and, 
hence the standard representation attached to the numbers of the
nodes only depends on the depth of the node in the tree, and on its rank on
its level. 

   If $a_k\ldots a_1a_0$ is the standard representation of $n$, say that
$n$ ends in $\alpha_1\alpha_0$.

   Following \cite{MMjucs}, call {\it continuator} of a node $n$ with
$a_k\ldots a_1a_0$ as its standard representation the node whose standard
representation is $a_k\ldots a_1a_000$. In the case of the standard Fibonacci
tree, the continuator of a node is always one of its sons and in that case, 
it is called the {\it preferred son}. In \cite{MMjucs}, we noticed that
in general the continuator of a node is not necessarily among the sons of the
node. There are trees in which for some nodes the continuator is not among the
sons and in the same trees, there are also nodes that contain two continuators 
of nodes among their sons. These properties are thoroughly studied in 
\cite{MMjucs} where we show that there is a continuous number of Fibonacci 
tree that possess the preferred son property for every node. If $n$ is a node, 
denote by $c(n)$ its continuator. We shall denote by $c^{-1}$ the converse 
operation: $c^{-1}(n)$ is the node with $a_k\ldots a_2$ as its standard 
representation where $a_k\ldots a_1a_0$ is the standard representation of $n$. 
Then, in the standard Fibonacci tree, the neighbours of a node $n$ are:
\vskip 5pt
{\leftskip 40pt\parindent 0pt
if $n$ is a 2-node which ends in 00:
\vskip 1pt\parindent 20pt
$c^{-1}(n)$, $c^{-1}(n)$$-$1, $c(n)$, $c(n)$+1, $c(n)$+2
\vskip 3pt\parindent 0pt
if $n$ is a 2-node which ends in 10:
\vskip 1pt\parindent 20pt
$c^{-1}(n)$+1, $c^{-1}(n)$, $c(n)$, $c(n)$+1, $c(n)$+2
\vskip 3pt\parindent 0pt
if $n$ is a 3-node:
\vskip 1pt\parindent 20pt
$c^{-1}(n)$, $c(n)$$-$1, $c(n)$, $c(n)$+1, $c(n)$+2
\par}
\vskip 5pt
\noindent
where the neighbours are indicated in the anticlockwise way.

   The path from the root to a given node is obtained by repeatedly applying
the $c^{-1}$ operation with a correction for 2-nodes that end in 10,
starting the process from the node and then taking the mirror of 
that list of nodes.

\subsection{Implementation issues}

   We consider now implementation issues.

   For that questions, we need to take the Fibonacci tree that will give us the
fastest algorithms that can solve the location problem. As an example, in the
case of the standard Fibonacci tree, the algorithms are linear, but as shown
in \cite{MMjucs}, for another tree, the algorithms are still linear, but a bit
faster.

   The reason of our choice is that in the case of the standard Fibonacci tree,
finding the {\it status} of a node $n$ is not immediate from the standard 
representation of $n$. The status is 2 when the last 2 digits are 10 and it is
3 when the last 2 digits are 01, but when it is 00, we have also to look at
the last two digits of the standard representation of node $n$$-$1:
if it is 10, $n$ is a 3-node, if it is 01, $n$ is a 2-node. The algorithm for
the neighbours is rather simple but it relies on the determination of the
status.

\vtop{
\vskip 10pt
\ligne{
\includegraphics[scale=0.7]{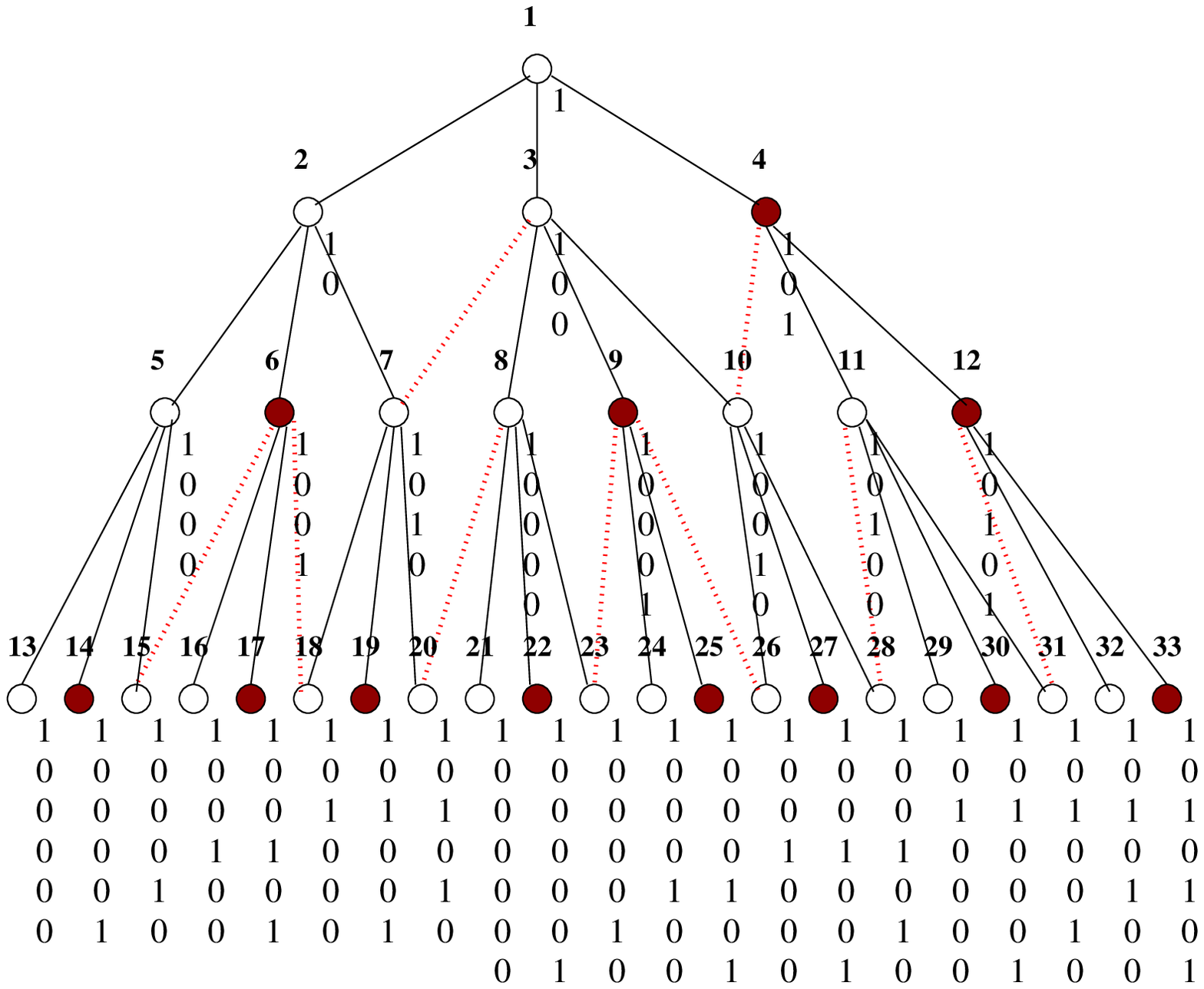}
\hfill}
\begin{fig}\label{otherfibotree}
\leurre
                              The best Fibonacci tree: 
\vskip0pt above a node: its number; below: its standard representation.
\end{fig}
\vskip 9pt
}

   As indicated in \cite{MMjucs} there is a Fibonacci tree such that its
2-nodes are exactly the nodes whose standard representation ends with 01.
This tree possesses the preferred son property and so, the algorithm for the
path and for the neighbours are linear too. For the path, the algorithm is
simpler as for the standard Fibonacci tree because the father of a node is 
always $c^{-1}$, the root and node~2 being excepted. For the neighbours, we 
have here the rules:
\vskip 9pt
{\leftskip 40pt\parindent 0pt
if $n$ is a 3-node that ends in 10:
\vskip 1pt\parindent 40pt
$c^{-1}(n)$, $c(n)$, $c(n)$+1, $c(n)$+2, $c^{-1}(n)$+1
\vskip 3pt\leftskip 60pt\parindent -20pt
if $n$ is a 2-node (it ends in 01) or if it is a $3$-node that ends in 00
for which $c^{-1}(n)$ is a 3-node:  
\vskip 1pt\parindent 20pt
$c^{-1}(n)$, $c(n)$$-$1, $c(n)$, $c(n)$+1, $c(n)$+2
\vskip 3pt\parindent -20pt
otherwise ($n$ is a 3-node that ends in 00 for which $c^{-1}(n)$ is a 2-node):
\vskip 1pt\parindent 20pt
$c^{-1}(n)$, $c^{-1}(n)$$-$1, $c(n)$, $c(n)$+1, $c(n)$+2
\par}
\vskip 5pt
We notice that there are indeed three possible configurations: for nodes in 01
and nodes in 00 whose the father is in 00 or 01, for nodes in 10 and for
nodes in 00 whose father is in 10, see Figure~\ref{otherfibotree}, below.

   Notice that besides the root for which the rule of the son is 
{\bf 3} $\rightarrow$ {\bf 3}$\,${\bf 3}$\,${\bf 2}, for all other nodes the 
rules are {\bf 2}$\rightarrow$ {\bf 3}$\,${\bf 2} for 2-nodes and 
{\bf 3} $\rightarrow$ {\bf 3}$\,${\bf 2}$\,${\bf 3}.

   According to this information, we can fix the format of the local rules
for cellular automata in the pentagrid. 

   First, we have to remark that the numbering of the nodes must be considered
as a hardware feature. In the Euclidean case, the same convention is done but
in that case the coordinate system is so evident that this fact is usually 
unnoticed. However, implementation programs have to fix such a system.

   Taking into account the considerations that we just developed, we can also
consider that the notions of father, of sons, of preferred son is also a
hardware feature. We express these properties when we say that any cell
{\it knows} which one is its father, what are the numbers of its neighbours,
what is its status and so on for similar questions.

   This allows us to consider only {\it states} of the automaton and to give a
uniform description. The format of the rules will be the following:

\vskip 5pt
\def\localrule fa #1 na #2 nb #3 nc #4 nd #5 old #6 nov #7 {%
\ligne{\hfill
\vtop{\offinterlineskip\parindent0pt\leftskip0pt\hsize=263.2pt
      \hrule height 0.3pt depth 0.3pt width \hsize
      \ligne{\vrule height 12pt depth 6pt width 0.6pt
             \hbox to 30pt{\hfill #1\hfill}
             \vrule height 12pt depth 6pt width 0.6pt
             \hbox to 30pt{\hfill #2\hfill}
             \vrule height 12pt depth 6pt width 0.6pt
             \hbox to 30pt{\hfill #3\hfill}
             \vrule height 12pt depth 6pt width 0.6pt
             \hbox to 30pt{\hfill #4\hfill}
             \vrule height 12pt depth 6pt width 0.6pt
             \hbox to 30pt{\hfill #5\hfill}
             \vrule height 12pt depth 6pt width 0.6pt
             \hbox to 30pt{\hfill #6\hfill}
             \vrule height 12pt depth 6pt width 0.6pt
             \hbox to 30pt{\hfill $\rightarrow$\hfill}
             \vrule height 12pt depth 6pt width 0.6pt
             \hbox to 30pt{\hfill #7\hfill}
             \vrule height 12pt depth 6pt width 0.6pt
             }
      \hrule height 0.3pt depth 0.3pt width \hsize
      }
\hfill}  
}
\localrule fa {\it fa} na {\it n$1$} nb {\it n$2$} nc {\it n$3$} 
           nd {\it n$4$} old {\it old} nov {\it new}  

\vskip 5pt
\vskip 5pt
\noindent
where $fa$ is the state of the father of the cell and $n1$, $n2$, $n3$
and $n4$ are the states of the other neighbours of the cell, listed in
an anticlockwise way starting from the father. Of course, {\it old} is the
current state of the cell and {\it new} is the state that it receives from
the application of the local rule.

\section{An application to tiling problems}

   The location technique that we give in sub-section 4.2. of this paper, 
also allows us to solve some tiling problems in the hyperbolic plane.

   In \cite{MMpavages}, one studies the tilings that can be generated by
replication of a single pentagon with coloured side, in such a way that
the colours of the sides do match and by using only displacements along the
sides of the tessellation. There are at most five colours that are denoted
by 1, 2, 3, 4 and 5. The question that is addressed by the paper is the 
following. If we give the number of colours and the assortment of the colours
on the sides of the initial pentagon, how many tilings can be generated under
the above restrictions? The answer is surprisingly not as trivial as it is
in the Euclidean case with a square.

   A basic remark is to notice that the displacements along the lines of the
tiling give for free the rotations that leave the tiling globally invariant.
This allows us to reduce the number of cases to be investigated and to consider
the assortments of colours up to circular permutations.
\vskip 5pt
\noindent
\begin{thm}\label{unipav}
{\rm (Margenstern)} $-$ 
The number of possible tilings with a 
single
pentagonal tile on the pentagrid with the assortments of $j$ colours,
$1\le j\le 5$ and satisfying the condition of displacements only along the
lines of the pentagrid
are given by 
Table~{\rm\ref{tabunipav}}.
\end{thm}
\vskip 10pt
\def\colori #1 #2 #3
          {\hrule height 0.3pt depth 0.3pt width \hsize
              \ligne{%
                     \vrule height 14pt depth 6pt width 0.6pt
                     \hbox to 15pt{\hskip 5pt #1\hfill}
                     \vrule height 14pt depth 6pt width 0.6pt
                     \hbox to 65pt{\hskip 6pt\tt #2\hfill}
                     \vrule height 14pt depth 6pt width 0.6pt
                     \hskip 5pt #3
                     \hfill
                     \vrule height 14pt depth 6pt width 0.6pt
                    }
           }

\vspace{-20pt}
\vtop{
\begin{tab}\label{tabunipav}
\leurre
Table of the results of theorem~{\rm\ref{unipav}}
\end{tab}
\vskip 0pt
\vspace{-10pt}
\ligne{\hfill\vtop{\parindent 0pt\leftskip 0pt\hsize=170pt
              \colori 5 {1 2 3 4 5} {no solution}
              \colori 4 {1 1 2 3 4} {2 solutions}
              \colori {} {1 2 1 3 4} {no solution}
              \colori 3 {1 1 1 2 3} {2$^{\aleph_0}$ solutions}
              \colori {} {1 1 2 1 3} {2$^{\aleph_0}$ solutions}
              \colori {} {1 1 2 2 3} {4 solutions}
              \colori {} {1 1 2 3 2} {2$^{\aleph_0}$ solutions}
              \hrule height 0.3pt depth 0.3pt width \hsize
                  }
       \hfill\vtop{\parindent 0pt\leftskip 0pt\hsize=170pt
              \colori {} {1 2 3 1 2} {no solution}
              \colori {} {1 2 3 1 3} {no solution}
              \colori 2 {1 1 1 1 2} {2$^{\aleph_0}$ solutions}
              \colori {} {1 1 1 2 2} {2$^{\aleph_0}$ solutions}
              \colori {} {1 1 2 1 2} {2$^{\aleph_0}$ solutions}
              \colori 1 {1 1 1 1 1} {1 solution}
              \hrule height 0.3pt depth 0.3pt width \hsize
             }
       \hfill}
\vskip 9pt
}

   In the case of the assortment \hskip-6pt\encadre{3pt}{\ 1 1 1 1 2\ }$\,$
where there is a continuous number of solutions, 
Figure~\ref{twounipav} illustrates two cases.

   Consider the picture~$(a)$ of that figure. 
It gives an example 
from that family of tilings. The construction starts from the standard 
Fibonacci tree, where all 2-nodes are on a same line starting from some initial
node which is a 3-node, among them the root. On that line, the side to the 
father is coloured with~2 for each two nodes. In the figure, we make use of 
the same colour for a line of 2-nodes that are connected in that way, the 
starting point of the line being a 3-node. We make use of 3 colours in order 
that the phenomenon should become clear.

   Now, il we randomly apply the rules used to define the succession of 2-nodes, namely
the rules \hbox{{\bf 2}$\rightarrow$ {\bf 2}$\,${\bf 3}} and
\hbox{{\bf 2}$\rightarrow$ {\bf 3}$\,${\bf 2}},
we obtain a continuous
family of tilings.

   We consider also another possibility which is given by the picture $(b)$ of
Figure~\ref{twounipav}. In this picture, pentagons are all linked together along their
single side that is coloured by~2. We can do a bit more: this association
of pairs, which gives rise to rectangular (non-regular) hexagons, can be
performed in such a way that each vertex is shared by exactly three hexagons.

\vskip 5pt
\vskip 5pt
\vtop{
\ligne{\hfill
\includegraphics[scale=0.5]{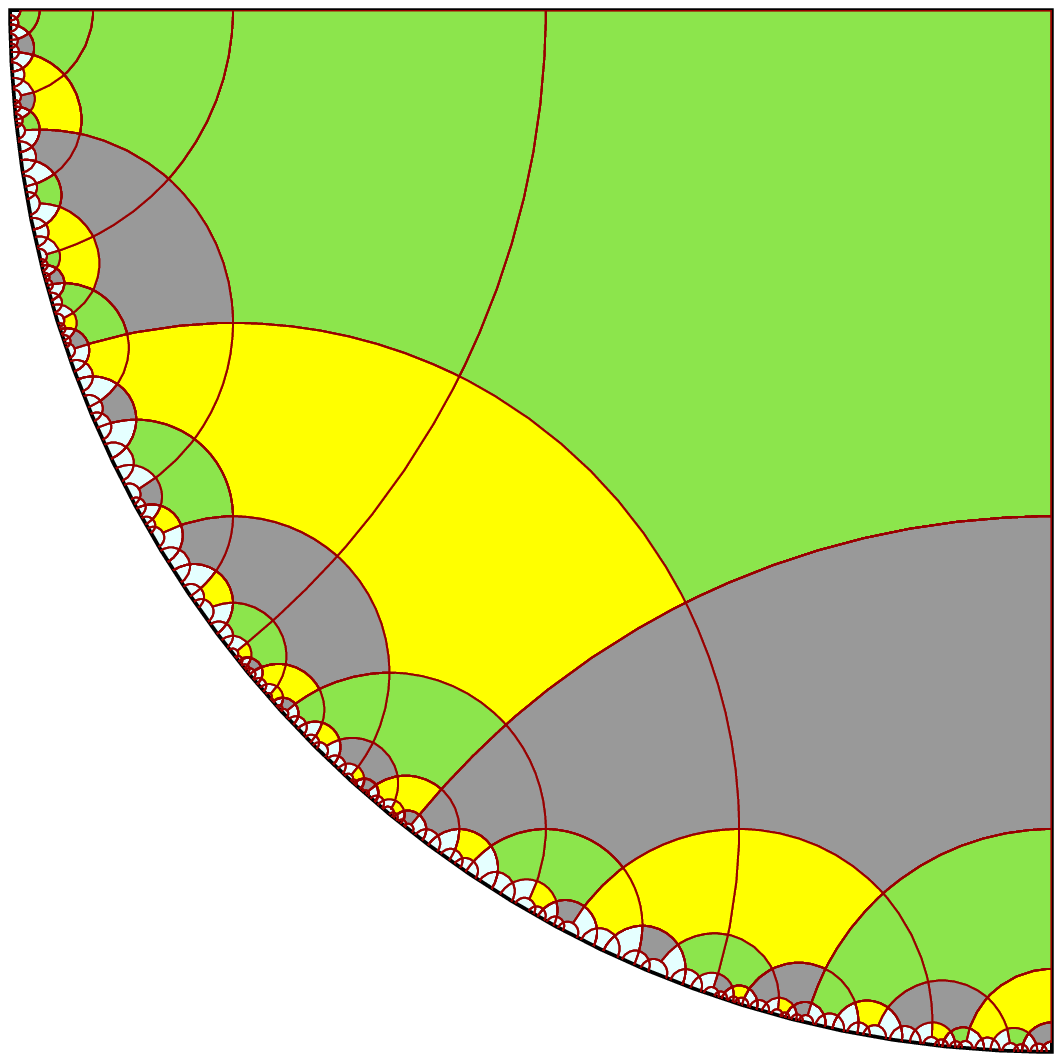}\hskip 6pt
\includegraphics[scale=0.5]{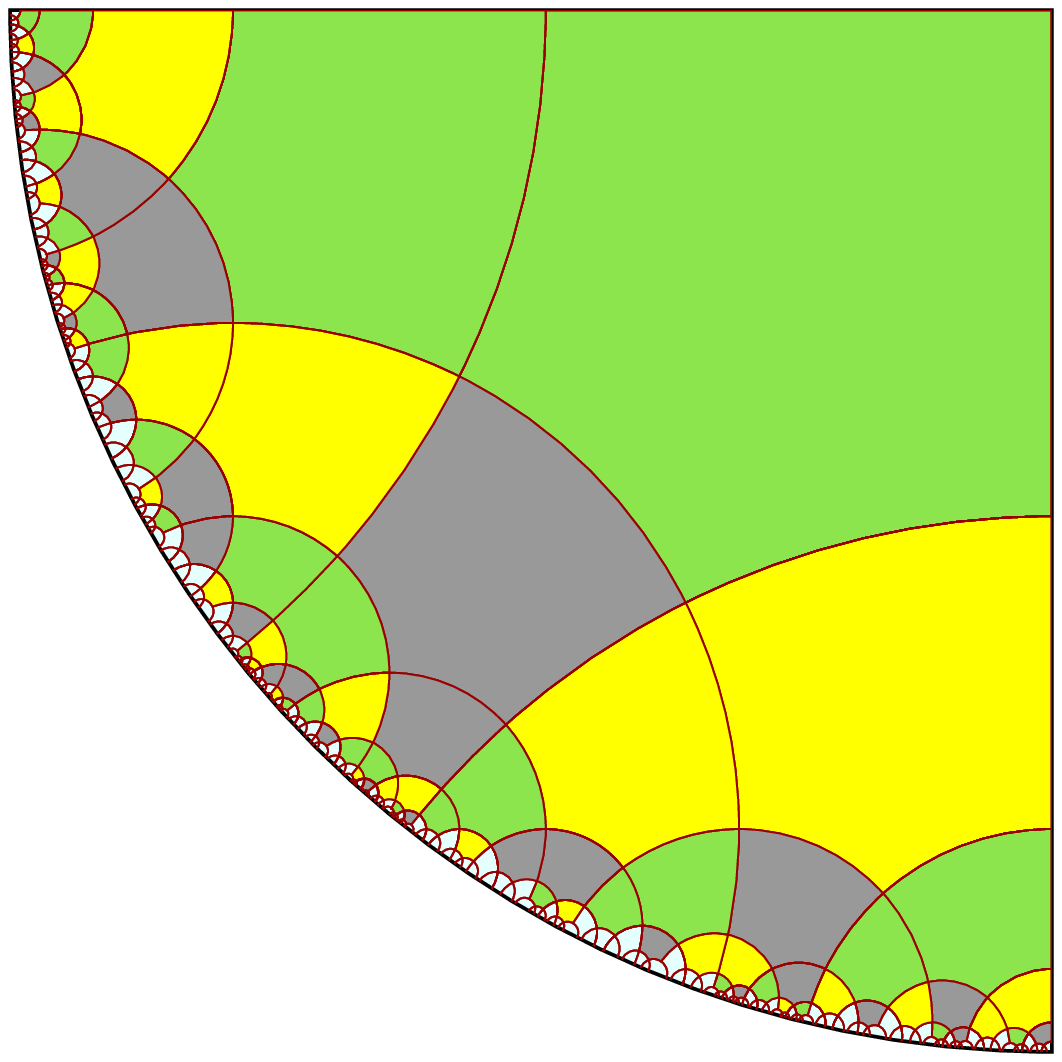}
\hfill}
\vskip -5pt
\ligne{\hfill
\PlacerEn {-250pt} {10pt} {$a$}
\PlacerEn {-50pt} {10pt} {$b$}
}
\vskip 5pt
\begin{fig}\label{twounipav}
Two tilings with colouring
\hskip-5pt\encadre{3pt}{\ $1$ $1$ $1$ $1$ $2$\ }$\,$. 
\end{fig}
\vskip 5pt
\vskip 5pt
}

   We turn now to the Cayley graph problem that we announced in section~3.

   In \cite{techrapp}, we show that the pentagrid is not the Cayley graph
of a sub-group of the isometries of the hyperbolic plane. The proof is by
cases: we show first that there should be at least one displacement. We easily 
rule out the case when all the sides of a pentagon represent displacements. 
And so, there is an even number of products of three reflections, they are
called {\it glides}, and at least one displacement. The different cases are
eliminated by considering a pair of independent vectors on one vertex of the
pentagon and on looking at the result of five isometries whose product should
be the identity, which is neither the case.

   However, the pentagrid is the Cayley graph of an abstract group. To see that
point, notice that if all generators that label the arcs from a vertex 
are involutions, {\it i.e.} $g^2=1$ for all those generators $g$, the existence
of the group boils down to the following tiling problem. We have at our disposal
four colours, 1, 2 3 and 4 and we have to colour the sides of all pentagons
of the grid in such a way that at any vertex, all the colours are present
on the sides that meet in the vertex. This can be performed by using the 
standard Fibonacci tree and proving by induction on the depth and on the rank
in a level that the process goes endlessly with success.

\section{About automaticity or not of groups on hyperbolic tilings}
\label{automaticity}

\subsection{The case of the pentagrid and of the heptagrid}

\begin{thm}\label{notauto} {\rm (Gasperin-Margenstern)}
Any group having the pentagrid or the heptagrid as a Cayley graph is not automatic.
\end{thm}

Proof of the theorem.
   The idea of the proof consists in looking at a family of closed paths in each of these
tilings which cannot be recognized by a finite automaton. We remind the reader that
a {\bf path} is a finite set of tiles \hbox{$T_0$,...,$T_n$} such that for 
\hbox{$i\in\{[0 \dots n-1]\}$}, $T_i$ and~$T_{i+1}$ share a common side. We say that
the {\bf length} of the path is~$n$ and that the path is {\bf closed} if
\hbox{$T_0=T_n$}.

   The family of paths we consider is the set of paths~$P_n$ defined as follows. 
The path starts from the leftmost son~$G$ of the root~$F$ of a Fibonacci tree~$\cal F$ of 
the tiling. Call~$\cal B$ the subtree of~$\cal F$ rooted at~$G$. Then, the follows the leftmost 
branch~$\lambda$ of~$\cal B$ until its level~$n$. There, it follows the level~$n$ 
until it meets the branch~$\rho$ of~$\cal B$ which is the rightmost branch of this tree. 
Then, when the path reaches~$\rho$, it goes back to the root of $\cal B$ by following~$\rho$.

   Assume that a group corresponding to the tiling is automatic. There is a finite automaton~$\cal A$
such that $\cal A$ recognizes the paths of the tiling which are closed and rejects those
which are not closed. Let~$Q$ be the set of states of~$\cal A$ and let \hbox{$N=\vert Q\vert$}, 
the number of elements of~$Q$. Let~$B$ be the finite set of symbols attached to each tile of 
the tiling and let \hbox{$r=\vert B\vert$}.
   
   It is plain that $\cal A$ accepts all~$P_n$'s. Consider a run of the automaton on $P_{r.N+2}$. 
Looking at the couples $(\alpha,q)$ read on the tiles of path which are on~$\lambda$, where
$\alpha$~is the symbol attached to the tile and $q$~is the state of~$\cal A$ when it reads this
tile, necessarily there are two tiles on the restriction of the path to~$\lambda$ on which
the couple~$(\alpha,q)$ are the same. Let $T_i$ and~$T_j$ be these tiles of~$P_{r.N+2}$. Then,
the subpath from $T_{i+1}$ up to~$T_j$ can be appended as many time as we wish in 
between~$T_j$ and~$T_{j+1}$. Let \hbox{$k=j$$-$$i$} and let~$m$ be the number of
iterations of the subpath from $T_{i+1}$ up to~$T_j$. Let $Q_{r.N+2,k,m}$ be the new path:
appending the subpath from $T_{i+1}$ up to~$T_j$ boils down to apply the translation~$\tau$
which transforms~$T_j$ into~$T_{i+km}$ to the tiles of~$P_{r.N+2}$ from~$T_j$ up to~$T_n$. 
The translation is correctly defined since by construction there exists a copy of the tree we move.
As $P_{r.N+2}$ goes back to~$G$ and as $\cal B$~is globally unchanged by 
the translation~$\tau$, $Q_{r.N+2,k,m}$ goes back to the image of~$G$
by~$\tau$: this image is not~$G$, the root of~$\cal B$. Now, the automaton also
recognizes $Q_{r.N+2,k,m}$, a contradiction as $Q_{r.N+2,k,m}$ is not closed.
\cqfd

\begin{cor}\label{nothyp}
There are infinitely many groups with Cayley graph corresponding to  tessellations of the hyperbolic plane
which are not hyperbolic.
\end{cor}

Proof.
   This argument can be repeated for all tessellations $\{p,4\}$ with~$p\geq5$ and
all tessellations $\{p,3\}$ with~$p\geq7$ as the tessellations $\{p,4\}$ 
and $\{p$+$2,3\}$ are generated by the same tree which generalizes the Fibonacci tree.
\cqfd

   Note that we have the following result, which may have a connection with Theorem~\ref{notauto}:

\begin{thm} \label{thm_non_cayley}{\rm (Margenstern, \cite{techrapp})}
The pentagrid is not the Cayley graph of a group of isometries leaving the pentagrid globally
invariant.
\end{thm}

   We repeat the proof of~\cite{techrapp} for the convenience of the reader. We use
Figure~\ref{non_cayley}, which is
a bit differrent from the figure used in~\cite{techrapp}.

\noindent
Proof. Assume that the pentagrid is the Cayley graph of a group~$G$ of isometries which
leave the pentagrid globally invariant: any pentagon of the tiling is transformed into
a pentagon of the tiling and the mapping is bijective. As we have four edges at any vertex,
$G$~has four generators~$g_1$, $g_2$, $g_3$ and~$g_4$.

   Consider a pentagon~$P_0$ whose vertices are denoted $A$, $B$, $C$, $D$ and~$E$, in this order
while clockwise turning around the tile, see Figure~\ref{non_cayley}. We may decide that~$P_0$ 
is the leading 
pentagon of a sector of the pentagrid. We give it number~1 and the pentagons~2, 3 and~4
are, by construction, those defined by the reflection of~$P_0$ in the sides~$ED$, $DC$ and~$CB$
respectively. In the figures, we indicate the numbering for a few pentagons which allows us
to easily derive all the numbers of the pentagons we shall consider in the proof.

\vtop{
\vskip 5pt
\ligne{\hfill
\includegraphics[scale=1]{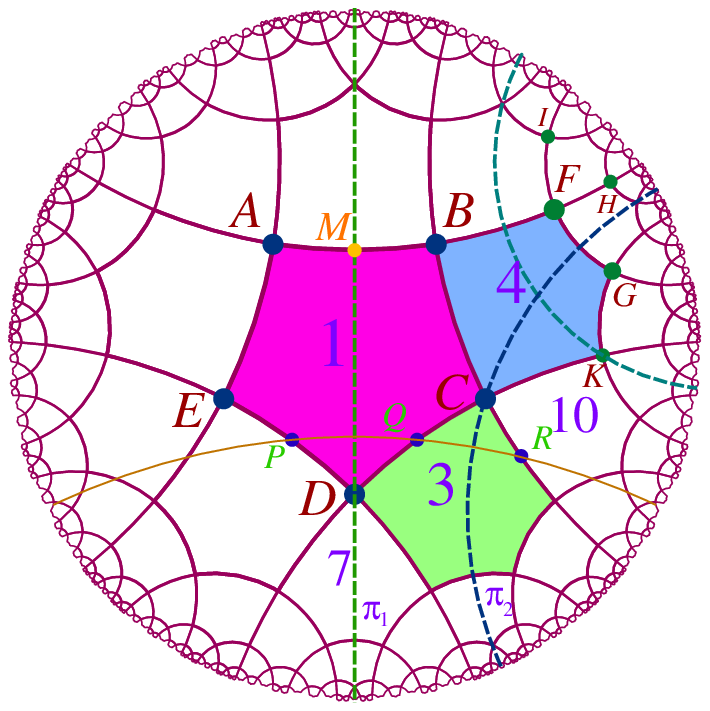}\hskip 6pt
\hfill}
\vskip 0pt
\vspace{-5pt}
\begin{fig}\label{non_cayley}
\leurre
The picture for the proof of Theorem~{\rm\ref{thm_non_cayley}}.
\end{fig}
}

We also know that, as in the Euclidean case, the products of reflections in lines which do not yield
the identity can be reduced two three kinds of transformations: one reflection, two reflections 
or three reflections. And so we shall speak of odd and even motions, the even motions
being exactly those which can be reduced to a product of two reflections in lines. The even
motions are characterized by the intersection property of the axes of the reflections which
define the motion: rotation, if the axes meet in the plane, ideal rotation, if the axes are
parallel, shift along the common perpendicular of the axes when they are non-secant.

We may assume that $g_1$~transforms~$A$ into~$B$
and we may assume that $g_1$ is an even motion: if all $g_i$'s are all odd motions, a product of
an odd number of them will remain odd and thus, cannot yield the identity.
Let~$F$ be the image of~$A$ under the reflection in~$B$. Let~$G$, $H$ and~$I$ be the other vertices
of the edges of pentagons which abut~$B$. If $g_1$ transforms~$F$ into~$B$, then the intersection
of the lines~$\pi_1$ and~$\pi_2$ the respective perpendicular of~$AB$ and~$FB$ tells us about the
nature of~$g_1$. Now, as $\pi_1$ and~$\pi_2$ have a common perpendicular, they are non-secant.
Moreover, as $g_1$~is a shift, the common perpendicular is the axis of the shift: the line which
is globally invariant under the shift. And so, $g_1$~is the shift along~$AB$ which transforms~$A$
into~$B$. We have the same conclusion, for a very similar reason if~$g_1$ transforms~$F$ into~$H$.

   Now, $g_1$ may also transform~$F$ into~$G$ or~$I$ instead of transforming it into~$B$ or~$H$.
We may look at the case of~$G$ as the case of~$I$ is very similar by reflection in~$AB$.
Again, we consider the intersection of~$\pi_1$ with~$\pi_2$ where $\pi_2$~is now the perpendicular
bisector of~$FG$. Let~$M$ be the midpoint of~$AB$. Clearly, the triangles $DBC$ and~$DAE$ are equal
from which we conclude that the triangle $ABD$~is isosceles with $AB$~as its basis and from this,
we easily conclude that $MD$~is also the perpendicular bisector of~$AB$ so that 
\hbox{$\pi_1=MD$}. From the equalities of angles at~$D$ we derive from this property, we
conclude that if $P$ and~$Q$ are the mid-points of~$ED$ and~$DC$ respectively, $\pi_1$~is also
the perpendicular bisector of~$PQ$. If we consider the perpendicular bisector of~$FG$, the just
performed argument shows that~$\pi_2$ passes through~$C$ and through~$N$ the mid-point of~$FG$.
It is also not difficult to prove that if~$R$ denotes the mid-point of the side of pentagon~10
which has~$C$ as a vertex and which is not shared by pentagon~4, then the triangle~$CQR$ is equal 
to the triangle~$DQP$. From
the previous study on pentagon~1 transported to pentagon~4, we know that the reflection of~$CQR$ 
in~$C$ is an isosceles triangle whose basis is the image of~$QR$ and the perpendicular bisector 
of the image of the basis is~$\pi_2$. Now, by the properties of the angles at~$C$ which are those 
of the angles at~$D$, we notice that~$\pi_2$ is globally invariant under the reflection in~$C$. 
Accordingly, $\pi_2$ is perpendicular to~$QR$. Now, from the equality of the triangles~$CQR$ 
and~$DQP$, it is not difficult to prove that the points $P$, $Q$ and~$R$ lie on the same line. 
Accordingly, $\pi_1$ and~$\pi_2$ also have a common perpendicular in this case and so $g_1$ is 
again a shift. Now, the line which is globally invariant under the shift~$g_1$ is the line~$PQR$.
Now, it is easy to show that the image of~$P$ under this shift, as it globally preserves the tiling, 
should be at least~$R$, as we have to consider the half-plane defined by~$AE$ which contains~$B$, 
but is in this case, the image of~$A$ is~$K$ and so, it cannot be~$B$. 

   And so, if $g_1$ is an even motion, it must be the shift along~$AB$
which transforms~$A$ into~$B$. Now, if all $g_i$'s are even motions,
they are shifts along sides of a pentagon which transform a vertex supported by the axis of the
shift to the other vertex of the side. Now, we can assume that we have five shifts $g_{i_1}$,
$g_{i_2}$, $g_{i_3}$, $g_{i_4}$ and~$g_{i_5}$ such that $g_{i_1}$ transforms~$A$ into~$B$,
$g_{i_2}$ transforms~$B$ into~$C$, $g_{i_3}$ transforms~$C$ into~$D$, $g_{i_4}$ transforms~$D$ 
into~$E$ and $g_{i_5}$ transforms~$E$ into~$A$. Now, considering the angle $(BA,BC)$, which is
inside pentagon~1,
it is not difficult to see that this angle is transformed into $(BC,BF)$ by
\hbox{$g_{i_1}\circ g_{i_2}\circ g_{i_3}\circ g_{i_4}\circ g_{i_5}$}, an angle which is outside
pentagon~1, so that this product cannot be the identity.

   The conclusion is that if
\hbox{$g_{i_1}\circ g_{i_2}\circ g_{i_3}\circ g_{i_4}\circ g_{i_5}=1$} there are
exactly two or exactly four distinct $g_i$'s that are odd motions of 
reflections in line. But the case of four distinct~$g_i$'s which are all odd motions has already
been ruled out. And so, two~$g_i$'s are odd motions and two~$g_i$'s are shifts.

   We may assume that~$g_1$ and~$g_2$ are the two generators which are odd motions.
And so $g_1$ is a reflection in a line or a product of three reflections
in lines. This latter situation is that of a glide and it can be shown that a glide
can be assumed to be in the form \hbox{$\rho_\ell\circ\rho_a\circ\rho_b$} where $a$, $b$~are 
non-secant and $\ell$ is the common perpendicular of~$a$ and~$b$. Consider $P_0$ a pentagon
and assume that $g_1$ is a glide transforming~$A$, a vertex of~$P_0$ into~$B$ the other vertex
of an edge~$c$ of~$P_0$. Now, consider \hbox{$\gamma=\rho_c\circ g_1$}. It is a product of an even
number of reflection of lines and so, $\gamma$~can be reduced to an even motion.
From what we have seen, it is the shift along~$c$ transforming~$A$ into~$B$. And so,
the glide is necessarily a reflection in an edge of a pentagon followed by a shift along 
the same edge. Note that in fact, the product of the reflection and the shift are here
commutative as they have the same axis.

   Note that if~$g_1$ is a single reflection in a line, as it transforms~$A$ into~$B$ it is
the reflection in~$\pi_1$. In all cases, we shall say that $AB$ is the side defining~$g_1$.

\newdimen\laboite\laboite=25pt
\newdimen\lelarge\lelarge=85pt
\def\laligne #1 #2 #3 {%
\hbox to \lelarge{
\hbox to \laboite{\hfill\tt#1\hfill}
\hbox to \laboite{\hfill\tt#2\hfill}
\hbox to \laboite{\hfill#3\hfill}
}
}

\vtop{
\vskip 7pt
\ligne{\hfill
\vtop{\leftskip 0pt\parindent 0pt\hsize=100pt
\laligne {$g_1$} {$g_2$} {}
\laligne g g 2
\laligne g r 4
\laligne r g 4
\laligne r r 2
}
\hskip 30pt
\vtop{\leftskip 0pt\parindent 0pt\hsize=100pt
\laligne {$g_1$} {$g_2$} {}
\laligne g g 4
\laligne g r 2
\laligne r g 2
\laligne r r 4
}
\hfill}
\begin{tab}\label{motions}
\leurre
Showing that 
\hbox{$g_{i_1}\circ g_{i_2}\circ g_{i_3}\circ g_{i_4}\circ g_{i_5}$} is not the identity. 
To left, the sides defining $g_1$ and~$g_2$ are contiguous. To right, the defining sides
are separated by one side of the pentagon. In all cases, the starting angle is~$1$.
\end{tab}
\vskip 7pt
}

Now, consider the second odd motion~$g_2$. There are two cases, depending whether the
side which defines $g_2$ is contiguous to that of~$g_1$ or not. If it is not the case,
we may assume that the side defining~$g_2$ is~$CD$ by symmetry of the pentagon. 
If~$g_2$ is defined by~$BC$, we have four cases, depending on the type of odd motions we have 
for~$g_1$ and~$g_2$. We also have four cases if $g_2$~is defined by~$CD$. We can indicate the
results by Table~\ref{motions} where we denoted the four angles around~$B$ as~1, 2, 3 and~4,
1 being the angle $(BA,BC)$ and the numbers being increasing while clockwise turning around the
vertex. The nature of the odd motion is indicated by the letters~$r$ and~$g$ for reflection in a
line and glide respectively. We always start from angle~1 and look at the result when applying
\hbox{$g_{i_1}\circ g_{i_2}\circ g_{i_3}\circ g_{i_4}\circ g_{i_5}$} where $g_{i_1}=g_1$.
We can see in Table~\ref{motions} that the result is never angle~1. Accordingly, 
Theorem~\ref{thm_non_cayley} is proved.
\cqfd

   However, if we remove the condition that the group should be a sub-group of the
group of isomorphisms which leaves the tiling globally invariant, then it is possible to represent the
pentagrid in this way as the following result states.

\begin{thm}\label{cayley_fantastic} {\rm (Margenstern, \cite{techrapp})}
The pentagrid is the Cayley graph of some abstract group.
\end{thm}

Proof.  Assume that there is a finitely generated group $G$ which is generated by four elements~$x$
such that $x^2=1$ for each of these elements and such that the pentagrid is the Cayley graph 
of~$G$.

   This is equivalent to the following condition. There is an assignment of each side of the
pentagons of the pentagrid onto a set of four colours~$S$ such that: any edge belonging to two 
neighbouring pentagons receive the same assignment in both pentagons and that at every vertex,
the sides abutting the vertex are coloured with all the colours of~$S$.

   We shall see that we can tile the pentagrid under these constraints to which we append a new one.
Denote the four colours by \hbox{\bf a$,$ b$,$ c} and~{\bf d}. While counter-clockwise turning 
around a tile, putting the colours into a word yields what we call a {\bf contour word} of the 
tile. Note that
the contour word is not unique: the different ones are obtained from each other by an appropriate
circular permutation. We define an $\alpha$-tile with 
\hbox{$\alpha\in\{$\bf a$,$b$,$c$\}$} as a tile whose contour word is
\hbox{$d\beta\gamma\beta\gamma$} with \hbox{$\{\alpha,\beta,\gamma\}=\{$\bf a$,$b$,$c$\}$}. We shall 
see that we can tile the pentagrid by $\alpha$-tiles only with 
\hbox{$\alpha\in\{$\bf a$,$b$,$c$\}$} and with
observing the constraint that all colours abut at each vertex.

   Let us fix a central tile~$C$. Let us put a $c$-tile at it. We shall denote its contour word
as \hbox{$d\alpha\beta\alpha\beta$} with \hbox{$\{\alpha,\beta\}=\{$\bf a$,$b$\}$}. Now, we can 
see that the five
roots of Fibonacci trees surrounding~$C$ can be denoted as follows:

\vskip 5pt
\ligne{\hfill
\vtop{\leftskip 0pt\parindent 0pt\hsize=100pt
\ligne{at $d$\hskip 20pt\hbox{$d\alpha\beta\alpha\beta$}\hfill}
\ligne{at $\alpha$\hskip 20pt\hbox{$\alpha\gamma\alpha\gamma d$}\hfill}
\ligne{at $\beta$\hskip 20pt\hbox{$\beta\gamma\beta\gamma d$}\hfill}
\ligne{at $\alpha$\hskip 20pt\hbox{$\alpha\gamma\alpha\gamma d$}\hfill}
\ligne{at $\beta$\hskip 20pt\hbox{$\beta\gamma\beta d\gamma$}\hfill}
}
\hfill}
\vskip 5pt

\noindent
where the first letter of the contour word corresponds to the side shared with~$C$. This dispatching
of the tiles is illustrated by Figure~\ref{central_cay}.

   Note that the nodes at~$\alpha$ and the first one at~$\beta$ can be considered of the same
type \hbox{$\alpha\beta\alpha\beta d$} as the letter~$d$ is always in the last position.
We can define the {\bf type} of a node~$\nu$ as the position of~$d$ among its neighbours, the father
of~$\nu$ being its neighbour~1 and numbering the other neighbours increasingly by counter-clockwise 
turning around~$\nu$. We can see that, around the central node, we can decide that three consecutive
nodes have type~5, one of the others having type~1 by construction: necessarily this
node is before the group of nodes with type~5 while counter-clockwise turning around the central
tile. It is not difficult to see that the other one has type~4.

\def\white #1 #2 #3 #4 #5 {%
\setbox210=\vbox{\leftskip 0pt\parindent 0pt\hsize=40pt
\ligne{\hfill#1\hfill}
\vskip 3pt
\hrule height 0.8pt depth 0pt width 40pt
}
\setbox211=\vtop{\leftskip 0pt\parindent 0pt\hsize=40pt
\hrule height 0pt depth 0.8pt width 40pt
\vskip 3pt
\ligne{\hfill#2\ #3\ #4\ #5\hfill}
}
\vbox{\box210\vskip-12pt\box211}
}

\def\black #1 #2 #3 #4 #5 {%
\setbox210=\vbox{\leftskip 0pt\parindent 0pt\hsize=30pt
\ligne{\hfill#2\ #1\hfill}
\vskip 3pt
\hrule height 0.8pt depth 0pt width 30pt
}
\setbox211=\vtop{\leftskip 0pt\parindent 0pt\hsize=30pt
\hrule height 0pt depth 0.8pt width 30pt
\vskip 3pt
\ligne{\hfill#3\ #4\ #5\hfill}
}
\vbox{\box210\vskip-12pt\box211}
}

\def\whitetypes #1 #2 #3 #4 #5 {%
\setbox210=\vbox{\leftskip 0pt\parindent 0pt\hsize=66pt
\ligne{\hfill#1\hfill}
\vskip 3pt
\hrule height 0.8pt depth 0pt width 66pt
}
\setbox211=\vtop{\leftskip 0pt\parindent 0pt\hsize=66pt
\hrule height 0pt depth 0.8pt width 66pt
\vskip 3pt
\ligne{\hfill\hbox to 12pt{\hfill#2\hfill}{\bf -} 
\hbox to 12pt{\hfill#3\hfill}
\hbox to 12pt{\hfill#4\hfill}$\bullet$
\hbox to 12pt{\hfill#5\hfill}
\hfill}
}
\vbox{\box210\vskip-12pt\box211}
}

\def\blacktypes #1 #2 #3 #4 {%
\setbox210=\vbox{\leftskip 0pt\parindent 0pt\hsize=66pt
\ligne{\hfill#1\hfill}
\vskip 3pt
\hrule height 0.8pt depth 0pt width 66pt
}
\setbox211=\vtop{\leftskip 0pt\parindent 0pt\hsize=66pt
\hrule height 0pt depth 0.8pt width 66pt
\vskip 3pt
\ligne{\hfill\hbox to 12pt{\hfill#2\hfill}{\bf -} 
\hbox to 12pt{\hfill#3\hfill}$\bullet$
\hbox to 12pt{\hfill#4\hfill}
\hfill}
}
\vbox{\box210\vskip-12pt\box211}
}

In order to avoid complicate figures on which not much is visible, we shall represent the colours
of a tile as follows: \white 1 2 3 4 5 for the white nodes and \black 1 2 3 4 5 for the black
ones, where in place of the numbers we put the colour associated to the side with this number.
We remind the reader that~1 is the side shared with the father and that the numbers are increasing
while counter-clockwise turning around the tile. We call this representation {\bf colour 
pattern}. Accordingly, the colour patterns of the neighbours of the central cell are:

\vskip 5pt
\ligne{\hfill
\white {\bf d} {$\alpha$} {$\beta$} {$\alpha$} {$\beta$}
\white {$\alpha$} {$\gamma$} {$\alpha$} {$\gamma$} {\bf d}
\white {$\beta$} {$\gamma$} {$\beta$} {$\gamma$} {\bf d}
\white {$\alpha$} {$\gamma$} {$\alpha$} {$\gamma$} {\bf d}
\white {$\beta$} {$\gamma$} {$\beta$} {\bf d} {$\gamma$} 
\white {\bf d} {$\alpha$} {$\beta$} {$\alpha$} {$\beta$},
\hfill}
\vskip 5pt
\noindent
where the types are~1, 5, 5, 5, 4 and 1 respectively, the first tile being repeated after the
last one in order to make the vertices in between four tiles all visible while turning around
the central cell. Here and in our sequel, we use $\alpha$, $\beta$ and~$\gamma$ as variables 
with the condition that \hbox{$\{\alpha,\beta,\gamma\}=\{$\bf a$,$ b$,$ c$\}$}.
Later, when black nodes also will appear, we shall append the mention of the status of the node
together with its type, as an example 1$_b$ for a black node of type~1.
If we consider two consecutive tiles, the two rightmost letters of the left-hand 
side tile, taking into account both lines of the colour pattern, and the two leftmost letters
of the right-hand side represent sides sharing a common vertex. Now we note that at each vertex
of tiles sharing a vertex, the four letters, 
\hbox{\bf a$,$ b$,$ c} and~{\bf d} are present. Now, looking at the above tiles, it can be seen
that when {\bf d} is on a side shared by a son of a node, for this son, 
{\bf d} appears on side~1. We shall see that for the other sons, we can decide that~{\bf d}
appears on side~2. Next, for nodes of type~2, we shall see that most sons we can decide that~{\bf d}
will appear on side~5. More rarely, types~4 and~3 will also appear.

\vtop{
\vskip 5pt
\ligne{\hfill
\includegraphics[scale=0.9]{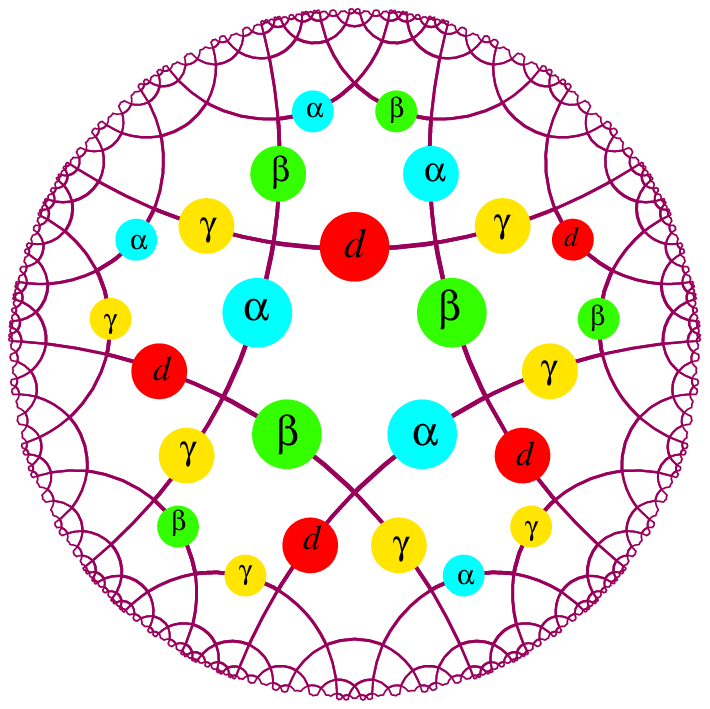}\hskip 6pt
\hfill}
\begin{fig}\label{central_cay}
\leurre
The distribution of $\alpha$-tiles around the central cell.
\end{fig}
}

   Let us now see the checking steps. They are structured in the same way: the configuration,
according to the pattern mentioned in Table~\ref{til_assort}, and then developed by using
the schemes above defined. There will be colours which can be defined by variables only with
conditions in order to preserve the presence of exactly the four colours at any vertex. We then
study the configuration for each son of the considered nodes and we shall summarize this in
a way again used in Table~\ref{til_assort} which gathers all results obtained in the discussion. 
\vskip 7pt
\noindent
$\underline{\hbox{\rm Configuration: \hbox{5 5}}}$
\vskip 5pt
For the right-hand side node, all its neighbours~$i$ with \hbox{$i\in\{2..5\}$}  can be assigned
the type~2.  Indeed, we have 
\white {$\alpha$} {$x$} {$\alpha$} {$x$} {\bf d}
\white {$\beta$} {$\gamma$} {$\beta$} {$\gamma$} {\bf d}, and now we can derive:
\vskip 5pt
\ligne{\hfill
\black {$x$} {\bf .} {\bf .} {\bf .} {$y$}
\white {$\alpha$} {\bf d} {$y$} {$\alpha$} {$y$}
\white {$x$} {\bf d} {$y$} {$x$} {$y$}
\black {$\gamma$} {\bf d} {$\alpha$} {$\gamma$} {$\alpha$}
\white {$\beta$} {\bf d} {$\alpha$} {$\beta$} {$\alpha$}
\white {$\gamma$} {\bf d} {$u$} {$\gamma$} {$u$}
\black {\bf .} {\bf d} {$v$} {\bf .} {$v$}
\hfill}
\vskip 5pt
\noindent
where \hbox{$\{x,y\}=\{\beta,\gamma\}$} and \hbox{$\{u,v\}=\{\alpha,\beta\}$}. 
As can be seen, all nodes but the first son are of type~2. For the first node, its type is 
in~\hbox{$\{2..4\}$}, depending on the left-hand side node with respect to
the node of type~5: \white {$\alpha$} {$x$} {$\alpha$} {$x$} {\bf d}{} $\,$.

We complete the situation of the neighbours of the central tile by considering the two sequences:
\hbox{5, 4} and also \hbox{4, 1, 5}. 

\vskip 7pt
\noindent
$\underline{\hbox{\rm Configuration: \hbox{5 4}}}$

We have:
\white {$\alpha$} {$\gamma$} {$\alpha$} {$\gamma$} {\bf d} 
\white {$\beta$} {$\gamma$} {$\beta$} {\bf d} {$\gamma$} {}, which gives us:
\vskip 5pt
\ligne{\hfill
\black {$\gamma$} {\bf .} {\bf .} {\bf .} {$\beta$} \hskip 5pt 
\white {$\alpha$} {\bf d} {$\beta$} {$\alpha$} {$\beta$} \hskip 5pt
\white {$\gamma$} {\bf d} {$\beta$} {$\gamma$} {$\beta$} \hskip 5pt
\black {$\gamma$} {\bf d} {$\alpha$} {$\gamma$} {$\alpha$} \hskip 5pt
\white {$\beta$} {\bf d} {$u$} {$\beta$} {$u$} \hskip 5pt
\white {\bf d} {$v$} {$\beta$} {$v$} {$\beta$} \hskip 5pt
\black {$\alpha$} {$\gamma$} {$\alpha$} {$\gamma$} {\bf d} \hskip 5pt
\hfill}
\vskip 5pt
\noindent
where \hbox{$\{u,v\}=\{\alpha,\gamma\}$}. 

We can summarize these relations by writing the types
of the nodes in the following way:
\hbox{\whitetypes 5 {$x$} 2 2 2 \whitetypes 4 2 2 1 5 }. The dash is put in between the black son
and the white ones of the same node, the~$\bullet$ is put between the last son of a node~$\nu$ and 
the first son of~$\nu$. Also note that the same black node appears twice in the above scheme:
once as the fifth neighbour of a node of type~5 and the second time as the first son of 
a node of type~4. 

\vskip 17pt
\noindent
$\underline{\hbox{\rm Configuration: \hbox{4 1 5}}}$
\vskip 5pt

We have:
\vskip 5pt
\ligne{\hfill 
\white {$\beta$} {$\gamma$} {$\beta$} {\bf d} {$\gamma$} {} \  
\white {\bf d} {$\alpha$} {$\beta$} {$\alpha$} {$\beta$} {} \
\white {$\beta$} {$\gamma$} {$\beta$} {\bf d} {$\gamma$} {}. 
\white {$\alpha$} {$\gamma$} {$\alpha$} {$\gamma$} {\bf d} 
\hfill}
\vskip 5pt
\noindent
Taking into account what we have found for~5 and~4,
we can now derive:
\vskip 5pt
\ligne{\hfill
\black {$\alpha$} {$\gamma$} {$\alpha$} {$\gamma$} {\bf d} \hskip 5pt
\white {$\beta$} {$\gamma$} {$\beta$} {$\gamma$} {\bf d} \hskip 5pt
\white {$\alpha$} {$\gamma$} {$\alpha$} {$\gamma$} {\bf d} \hskip 5pt
\black {$\gamma$} {$\beta$} {$\gamma$} {\bf d} {$\beta$} \hskip 5pt
\white {$\alpha$} {\bf d} {$\beta$} {$\alpha$} {$\beta$} \hskip 5pt
\white {$\gamma$} {\bf d} {$x$} {$\gamma$} {$x$} \hskip 5pt
\black {\bf .} {\bf d} {$y$} {\bf d} {$\beta$} \hskip 5pt
\hfill}
\vskip 5pt
\noindent
with \hbox{$\{x,y\}=\{\alpha,\beta\}$}. This information can be rewritten as:
\vskip 5pt
\ligne{\hfill\whitetypes 4 {$x$} 2 1 5 \whitetypes 1 5 5 5 4 \whitetypes 5 4 2 2 2 .\hfill} 
\vskip 5pt
\noindent 
Note the node of
type~4 which appears here is a black node.
\vskip 10pt
   At this step, several new patterns have to be analyzed: \hbox{2 2 2$_b$}, \hbox{2$_b$ 2 1}, 
\hbox{2 1 5$_b$}, \hbox{5$_b$ 5}
and \hbox{5 4$_b$ 2}.

\vskip 17pt
\noindent
$\underline{\hbox{\rm Configuration: \hbox{2 2 2$_b$}}}$

Accordingly, 
we consider the following situation:
\hbox{\white {$\alpha$} {\bf d} {$\gamma$} {$\alpha$} {$\gamma$}
\white {$\beta$} {\bf d} {$\gamma$} {$\beta$} {$\gamma$} 
\black {\bf d} {$\alpha$} {\bf .} {\bf .} {\bf .} }. It is not difficult to check that this time
we obtain:
\vskip 5pt
\ligne{\hfill
\black {\bf d} {\bf .} {$z$} {\bf .} {$z$}
\white {$\gamma$} {$y$} {$\gamma$} {$y$} {\bf d}
\white {$\alpha$} {$\beta$} {$\alpha$} {$\beta$} {\bf d}
\black {\bf d} {$\gamma$} {$\beta$} {$\gamma$} {$\beta$}
\white {$\gamma$} {$\alpha$} {$\gamma$} {$\alpha$} {\bf d}
\white {$\beta$} {$\alpha$} {$\beta$} {$\alpha$} {\bf d}
\black {$\alpha$} {$\gamma$} {$\alpha$} {$\gamma$} {\bf d}
\white {$x$} {$u$} {$x$} {$u$} {\bf d}
\hfill}
\vskip 5pt
\noindent
with \hbox{$\{y,z\}=\{\alpha,\beta\}$} and
\hbox{$\{u,x\}=\{\beta,\gamma\}$}. We can summarize these results as
\ligne{\hfill
\vtop{\leftskip 0pt\parindent 0pt
\vskip 2pt
\hbox{\whitetypes 2 1 5 5 1 \whitetypes 2 1 5 5 5 \blacktypes 2 5 5 {$x$} }.
\vskip 5pt
}
\hfill}

\vskip 17pt
\noindent
$\underline{\hbox{\rm Configuration: \hbox{2$_b$ 2 1}}}$

The nodes are given by the following
patterns: 
\hbox{\black {$\alpha$} {\bf d} {$\gamma$} {$\alpha$} {$\gamma$} , 
\white {$\beta$} {\bf d} {$u$} {$\beta$} {$u$} } and
\white {\bf d} {$x$} {$v$} {$x$} {$v$} ,
with \hbox{$\{u,x\}=\{\alpha,\gamma\}$} and
\hbox{$\{v,w,x\}=\{\alpha,\beta,\gamma\}$}. Using what we have seen for~1 with
\hbox{4 1 5}, the computation yields:
\vskip 5pt
\ligne{\hfill
\black {$\gamma$} {\bf .} {$\gamma$} {\bf .} {\bf d}
\white {$\alpha$} {$\beta$} {$\alpha$} {$\gamma$} {\bf d}
\black {\bf d} {$\gamma$} {$\beta$} {$\gamma$} {$\beta$}
\white {$u$} {$x$} {$u$} {$x$} {\bf d}
\white {$\beta$} {$x$} {$\beta$} {$x$} {\bf d}
\black {$x$} {$u$} {$x$} {$u$} {\bf d}
\white {$v$} {$w$} {$v$} {$w$} {\bf d}
\hfill}
\vskip 5pt
\noindent
with
\hbox{$\{u,x\}=\{\alpha,\gamma\}$} and  \hbox{$\{v,w,x\}=\{\alpha,\beta,\gamma\}$}. A
possible assignment of the variables is \hbox{$x=\alpha$}, \hbox{$v=\beta$} and
\hbox{$u=w=\gamma$}. This can be summarized as:
\vskip 5pt
\ligne{\hfill
\blacktypes 2 5 5 1 \whitetypes 2 1 5 5 5 \whitetypes 1 5 5 5 {$x$}
\hfill}
\vskip 5pt
\noindent
Note, that in \hbox{4 1 5}, the last node of~1 is~4 and here, it is not determined.

\vskip 17pt
\noindent
$\underline{\hbox{\rm Configuration: \hbox{2 1 5$_b$}}}$

This means that the nodes we consider
are \hbox{\white {$\alpha$} {\bf d} {$x$} {$\alpha$} {$x$}
\white {\bf d} {$y$} {$z$} {$y$} {$z$} } and
\black {$t$} {$\beta$} {$t$} {$\beta$} {\bf d} {}, with
\hbox{$\{x,y\}=\{\beta,\gamma\}$} and  \hbox{$\{t,z\}=\{\alpha,\gamma\}$}. As an example,
\hbox{$y=t=\gamma$}, \hbox{$x=\beta$} and \hbox{$z=\alpha$} are possible choices.
Without repeating previous results for the sons of a node of type~2, computations yield:
\vskip 5pt
\ligne{\hfill
\white {$\alpha$} {$\beta$} {$\alpha$} {$\beta$} {\bf d} \hskip 5pt
\black {$\beta$} {$\gamma$} {$\beta$} {$\gamma$} {\bf d} \hskip 5pt
\white {$z$} {$w$} {$z$} {$w$} {\bf d} \hskip 5pt
\white {$\beta$} {$w$} {$\beta$} {$w$} {\bf d} \hskip 5pt
\black {$w$} {$z$} {$w$} {$z$} {\bf d} 
\white {$\beta$} {$z$} {$\beta$} {\bf d} {$z$} \hskip 5pt
\hfill}
\vskip 5pt
\noindent
with
\hbox{$\{u,v\}=\{\alpha,\beta\}$} and  \hbox{$\{w,z\}=\{\alpha,\gamma\}$}.
This gives us:
\vskip 5pt
\ligne{\hfill
\whitetypes 2 1 5 5 5 \whitetypes 1 5 5 5 5 \blacktypes 5 5 4 2
\hfill}
\vskip 5pt

\vskip 17pt
\noindent
$\underline{\hbox{\rm Configuration: \hbox{5$_b$ 5}}}$

It is very similar to the case 5~5. We have:
\hbox{\black {$\beta$} {$x$} {$\beta$} {$x$} {\bf d}
\white {$\alpha$} {$\gamma$} {$\alpha$} {$\gamma$} {\bf d} }, with \hbox{$x\in\{\alpha,\gamma\}$}.
Computations give us:
\vskip 5pt
\ligne{\hfill
\black {$\beta$} {\bf .} {\bf .} {\bf .} {$y$} \hskip 5pt
\white {$x$} {\bf d} {$y$} {$x$} {$y$} \hskip 5pt
\black {$\gamma$} {\bf d} {$\beta$} {$\gamma$} {$\beta$} \hskip 5pt
\white {$\alpha$} {\bf d} {$\beta$} {$\alpha$} {$\beta$} \hskip 5pt
\white {$\gamma$} {\bf d} {$u$} {$\gamma$} {$u$} \hskip 5pt
\black {\bf .} {\bf d} {$v$} {\bf .} {\bf .} \hskip 5pt
\hfill}
\vskip 5pt
\noindent
with \hbox{$\{x,y\}=\{\alpha,\gamma\}$} and \hbox{$\{u,v\}=\{\alpha,\beta\}$}. This can be
summarized by: 
\vskip 5pt
\ligne{\hfill\blacktypes 5 {$x$} 2 2 \whitetypes 5 2 2 2 2 ,\hfill}
\vskip 5pt
\noindent
which is very close to whet we obtained for 5 5.

\vskip 17pt
\noindent
$\underline{\hbox{\rm Configuration: \hbox{5 4$_b$ 2}}}$

We have: \hbox{\white {$\alpha$} {$y$} {$\alpha$} {$y$} {\bf d} {}
\black {$\gamma$} {$\beta$} {$\gamma$} {\bf d} {$\beta$}  
\white {$\alpha$} {\bf d} {$x$} {$\alpha$} {$x$} .}
This give us:
\vskip 5pt
\ligne{\hfill
\white {$\alpha$} {\bf d} {$z$} {$y$} {$z$}
\white {$y$} {\bf d} {$h$} {$y$} {$h$}
\black {$\gamma$} {\bf d} {$k$} {$\gamma$} {$k$}
\white {\bf d} {$t$} {$u$} {$t$} {$u$}
\black {\bf d} {$\beta$} {$v$} {$\beta$} {$v$}
\white {$x$} {$w$} {$x$} {$w$} {\bf d}
\white {$\alpha$} {$s$} {$\alpha$} {$s$} {\bf d}
\black {\bf .} {$x$} {$s$} {\bf .} {\bf .}
\hfill}
\vskip 5pt
\noindent
with \hbox{$\{y,z\}=\{s,x\}=\{\beta,\gamma\}$}, \hbox{$\{k,t\}=\{\alpha,\beta\}$},
\hbox{$\{h,k,y\}=\{v,w,x\}=\{\alpha,\beta,\gamma\}$} and 
\hbox{$\{u,v\}=\{\alpha,\gamma\}$}.
A possible assortment of these variables is obtained by setting \hbox{$k=u=w=\alpha$}, 
\hbox{$t=x=y=\beta$}
and \hbox{$h=s=v=z=\gamma$}. We can summarized the results as:
\vskip 5pt
\ligne{\hfill
\whitetypes 5 {$x$} 2 2 2 \blacktypes 4 2 1 1 \whitetypes 2 1 5 5 {$y$}
\hfill}
\vskip 5pt
\vskip 5pt
   At this new step, we again have new patterns to study: 
\hbox{5 5$_b$}, \hbox{2$_b$ 1 1$_b$},
\hbox{5$_b$ 4 2$_b$}, \hbox{5 1$_b$ 5} and \hbox{1 1$_b$ 5}

\vskip 17pt
\noindent
$\underline{\hbox{\rm Configuration: \hbox{5 5$_b$}}}$

We shall find a situation very similar to
\hbox{5 5} and \hbox{5$_b$ 5}. We have:
\hbox{\white {$\alpha$} {$x$} {$\alpha$} {$x$} {\bf d} 
\black {$\gamma$} {$\beta$} {$\gamma$} {$\beta$} {\bf d} }, with
\hbox{$x\in\{\beta,\gamma\}$}. Computations give us:
\vskip 5pt
\ligne{\hfill
\black {$x$} {\bf .} {\bf .} {\bf .} {$y$} 
\white {$\alpha$} {\bf d} {$y$} {$\alpha$} {$y$} 
\white {$x$} {\bf d} {$y$} {$x$} {$y$} 
\black {$\gamma$} {\bf d} {$\alpha$} {$\gamma$} {$\alpha$} 
\white {$\beta$} {\bf d} {$u$} {$\beta$} {$u$} 
\black {\bf .} {\bf d} {$v$} {\bf .} {$v$} 
\hfill}
\vskip 5pt
\noindent
with now \hbox{$\{x,y\}=\{\beta,\gamma\}$} and \hbox{$\{u,v\}=\{\alpha,\gamma\}$}.
A possible assignment is \hbox{$u=\alpha$}, \hbox{$x=\beta$} and \hbox{$v=y=\gamma$}.
The result can be summarized as:
\vskip 5pt
\ligne{\hfill
\whitetypes 5 {$x$} 2 2 2 \blacktypes 5 2 2 2 ,
\hfill}
\vskip 5pt
\noindent
which looks very much to the results obtained with \hbox{5 5} and \hbox{5$_b$ 5}, see
Table~\ref{til_assort}.

\vskip 17pt
\noindent
$\underline{\hbox{\rm Configuration: \hbox{2$_b$ 1 1$_b$}}}$

We have: \hbox{\black {$\alpha$} {\bf d} {$\beta$} {$\alpha$} {$\beta$} 
\white {\bf d} {$\gamma$} {$y$} {$\gamma$} {$y$} 
\black {\bf d} {$x$} {$z$} {$x$} {$z$} ,}
with \hbox{$\{x,y,z\}=\{\alpha,\beta,\gamma\}$}. The computations yield:
\vskip 5pt
\ligne{\hfill
\black {$\beta$} {\bf .} {\bf .} {$\beta$} {\bf d} 
\white {$\alpha$} {$\gamma$} {$w$} {$\gamma$} {\bf d} 
\black {$\gamma$} {$\beta$} {$\gamma$} {$\beta$} {\bf d} 
\white {$y$} {$u$} {$y$} {$u$} {\bf d} 
\white {$\gamma$} {$u$} {$\gamma$} {$u$} {\bf d} 
\black {$z$} {$y$} {$z$} {$y$} {\bf d} 
\white {$x$} {$y$} {$x$} {$y$} {\bf d} 
\black {\bf .} {$z$} {$y$} {\bf .} {\bf .} 
\hfill}
\vskip 5pt
\noindent
with also \hbox{$\{u,y\}=\{\alpha,\beta\}$} \hbox{$w\in\{\alpha,\beta\}$}. A possible assignment 
of the variables is \hbox{$x=u=\alpha$}, \hbox{$w,y=\beta$} and \hbox{$z=\gamma$}.
The result can be summarized by:
\vskip 5pt
\ligne{\hfill
\blacktypes 2 5 5 5 \whitetypes 1 5 5 5 5 \blacktypes 1 5 5 {$x$}
\hfill}
\vskip 5pt

\vskip 17pt
\noindent
$\underline{\hbox{\rm Configuration: \hbox{5$_b$ 4 2$_b$}}}$

We have: \hbox{
\black {$\beta$} {$\alpha$} {$\beta$} {$\alpha$} {\bf d} 
\white {$x$} {$y$} {$x$} {\bf d} {$y$} 
\black {$u$} {\bf d} {$\beta$} {$u$} {$\beta$} 
}
where \hbox{$\{u,x,y\}=\{\alpha,\gamma\}$}. Computations yield:
\vskip 5pt
\ligne{\hfill
\black {$\beta$} {\bf .} {\bf .} {\bf .} {$\gamma$} 
\white {$\alpha$} {\bf d} {$\gamma$} {$\alpha$} {$\gamma$} 
\black {$y$} {\bf d} {$\beta$} {$y$} {$\beta$} 
\white {$x$} {\bf d} {$z$} {$x$} {$z$} 
\white {\bf d} {$t$} {$w$} {$t$} {$w$} 
\black {$\beta$} {$y$} {$\beta$} {$y$} {\bf d} 
\white {$u$} {$v$} {$u$} {$v$} {\bf d} 
\black {\bf .} {$\beta$} {$v$} {\bf .} {\bf .} 
\hfill}
\vskip 5pt
\noindent
with \hbox{$\{x,y\}=\{\alpha,\gamma\}$}, \hbox{$\{u,v\}=\{\alpha,\gamma\}$},
\hbox{$\{t,x,z\}=\{\alpha,\beta,\gamma\}$} and \hbox{$\{w,y\}=\{\alpha,\gamma\}$}.
A possible assignment of the variables is \hbox{$u=w=x=\alpha$}, \hbox{$t=\beta$}
and \hbox{$v=y=\gamma$}. We can summarize the result by:
\vskip 5pt
\ligne{\hfill
\blacktypes 5 {$x$} 2 2 \whitetypes 4 2 2 1 5 \blacktypes 2 5 5 {$y$}
\hfill}

\vskip 17pt
\noindent
$\underline{\hbox{\rm Configuration: \hbox{5 1$_b$ 5}}}$

We have: \hbox{
\white {$\alpha$} {$x$} {$\alpha$} {$x$} {\bf d} 
\black {\bf d} {$\beta$} {$\gamma$} {$\beta$} {$\gamma$} 
\white {$r$} {$s$} {$r$} {$s$} {\bf d} ,}
with \hbox{$x\in\{\beta,\gamma\}$} and \hbox{$\{r,s\}=\{\alpha,\beta\}$}. Computations
give us:
\vskip 5pt
\ligne{\hfill
\black {$x$} {\bf .} {\bf .} {\bf .} {$y$} 
\white {$\alpha$} {\bf d} {$y$} {$\alpha$} {$y$} 
\white {$x$} {\bf d} {$y$} {$x$} {$y$} 
\black {$\gamma$} {\bf d} {$\alpha$} {$\gamma$} {$\alpha$} 
\white {$\beta$} {\bf d} {$\alpha$} {$\beta$} {$\alpha$} 
\black {$s$} {$\gamma$} {\bf d} {$s$} {$\gamma$} 
\white {$r$} {\bf d} {$\gamma$} {$r$} {$\gamma$} 
\white {$s$} {\bf d} {$t$} {$s$} {$t$}
\hfill}
\vskip 5pt
\noindent
with \hbox{$\{x,y\}=\{\beta,\gamma\}$}, \hbox{$\{u,x\}=\{\beta,\gamma\}$},
\hbox{$\{r,s\}=\{\alpha,\beta\}$} and \hbox{$\{s,t,v\}=\{\alpha,\beta,\gamma\}$}.
A possible assignment of the variables is \hbox{$r=t=\alpha$},
\hbox{$s=x=\beta$} and \hbox{$u=v=y=\gamma$}. We note that we have a new pattern which is
of type~3 and it is a black node. In fact, it behaves as a white node of type~2, as shall be
seen later. The results can be summarized as:
\vskip 5pt
\ligne{\hfill
\whitetypes 5 {$x$} 2 2 2 \blacktypes 1 2 2 3 \whitetypes 5 3 2 2 2 
\hfill}

\vskip 17pt
\noindent
$\underline{\hbox{\rm Configuration: \hbox{1 1$_b$ 5}}}$

We have: 
\hbox{
\white {\bf d} {$x$} {$\beta$} {$x$} {$\beta$} 
\black {\bf d} {$\alpha$} {$\gamma$} {$\alpha$} {$\gamma$} 
\white {$\alpha$} {$\beta$} {$\alpha$} {$\beta$} {\bf d} ,
\hfill}
with \hbox{$x\in\{\alpha,\gamma\}$}. Computations yield:
\vskip 5pt
\ligne{\hfill
\black {$\alpha$} {$u$} {$\alpha$} {$u$} {\bf d} 
\white {$\beta$} {$\gamma$} {$\beta$} {$\gamma$} {\bf d} 
\white {$\alpha$} {$\gamma$} {$\alpha$} {$\gamma$} {\bf d} 
\black {$\gamma$} {$\beta$} {$\gamma$} {$\beta$} {\bf d} 
\white {$\alpha$} {$\beta$} {$\alpha$} {$\beta$} {\bf d} 
\black {$\beta$} {$\gamma$} {$\beta$} {\bf d} {$\gamma$}
\white {$\alpha$} {\bf d} {$\gamma$} {$\alpha$} {$\gamma$} 
\white {$\beta$} {\bf d} {$u$} {$\beta$} {$u$} 
\hfill}
\vskip 5pt
\noindent
where we find that $x=\alpha$ and we need \hbox{$\{u,v\}=\{\alpha,\gamma\}$}.
We can summarize the result as:
\vskip 5pt
\ligne{\hfill
\whitetypes 1 5 5 5 5 \blacktypes 1 5 5 4 \whitetypes 5 4 2 2 2 
\hfill}
\vskip 5pt
\vskip 5pt
\vskip 5pt
   At this step, the single new configuration is \hbox{2 3$_b$ 2}. 
\vskip 17pt
\noindent
$\underline{\hbox{\rm Configuration: \hbox{2 3$_b$ 2}}}$

We have: \hbox{
\white {$\alpha$} {\bf d} {$\gamma$} {$\alpha$} {$\gamma$}
\black {$x$} {$\beta$} {\bf d} {$x$} {$\beta$}
\white {$y$} {\bf d} {$z$} {$y$} {$z$} }
with \hbox{$\{x,y\}=\{\alpha,\gamma\}$} and \hbox{$z\not=y$}. Computations yield:
\vskip 5pt
\ligne{\hfill
\black {\bf d} {\bf .} {$s$} {\bf .} {$s$}
\white {$\gamma$} {$r$} {$\gamma$} {$r$} {\bf d}
\white {$\alpha$} {$\beta$} {$\gamma$} {$\beta$} {\bf d}
\black {\bf d} {$\gamma$} {$\beta$} {$\gamma$} {$\beta$}
\white {$x$} {$y$} {$x$} {$y$} {\bf d}
\black {\bf d} {$\beta$} {$y$} {$\beta$} {$y$}
\white {$z$} {$u$} {$z$} {$u$} {\bf d}
\white {$y$} {$u$} {$y$} {$u$} {\bf d}
\hfill}
\vskip 5pt
\noindent
with \hbox{$\{r,s\}=\{\alpha,\beta\}$}, \hbox{$\{x,y\}=\{\alpha,\gamma\}$}
and \hbox{$\{u,y,z\}=\{\alpha,\beta,\gamma\}$}. A possible assignment of the 
variables is \hbox{$r=u=x=\alpha$}, \hbox{$s=z=\beta$} and \hbox{$y=\gamma$}.
The result can be summarized by:
\vskip 5pt
\ligne{\hfill
\whitetypes 2 1 5 5 1 \blacktypes 3 1 5 1 \whitetypes 2 1 5 5 {$x$}
\hfill}
\vskip 5pt
We can note that there is no new configuration and that the types generated in this case
are very similar to what is produced by a sequence of consecutive nodes of type~2.
\vskip 10pt
   In Table~\ref{til_assort}, for each pattern, we indicate the configuration of the sons
of each node of the pattern. If the nodes belonging to the pattern belong to level~$n$ of the
tree, all the nodes appearing as a son of a node in the pattern are at the level~$n$+1. The
table numbers each pattern and, in the last columns, for each pattern~$p$, it indicates the 
number of the patterns raised by~$p$. In this way, we can see that as no new pattern appear after
the 15 patterns of the table, the induction can prove that if the tiling is possible at the
central cell and at the level of its immediate neighbours around it, the tiling is possible
for all the next levels. This is strengthened by the fact that the colours of the sons are not
completely fixed, which allows to make the choices which match with the constraints. Accordingly,
this proves Theorem~\ref{cayley_fantastic}.
\def\tabulure #1 #2 #3 #4%
{
\ligne{\hbox to 20pt{\footnotesize\tt#1\hfill}\hbox to 50pt{#2\hfill}
\hbox to 230pt{#3\hfill}\hbox to 30pt{\hfill\footnotesize\tt#4}\hfill
}
\vskip 5pt
}
\vtop{
\vspace{-5pt}
\begin{tab}\label{til_assort}
\leurre
The types of the sons of nodes with given types. The first column indicates the number of
the pattern. On each line, the last column indicates the numbers of the patterns involved by 
the pattern described in the line.
\end{tab}
\ligne{\hfill
\vspace{-6pt}
\vtop{\parindent 0pt\leftskip 0pt\hsize=335pt
\tabulure {} {pattern} {patterns of the sons} {}
\vskip 5pt
\tabulure 1 {5 5} {\whitetypes 5 {$x$} 2 2 2 \whitetypes 5 2 2 2 2 } 6
\tabulure 2 {5 5$_b$} {\whitetypes 5 {$x$} 2 2 2 \blacktypes 5 2 2 2 } 6
\tabulure 3 {5$_b$ 5} {\blacktypes 5 {$x$} 2 2 \whitetypes 5 2 2 2 2 } 6
\tabulure 4 {5 4} {\whitetypes 5 {$x$} 2 2 2 \whitetypes 4 2 2 1 5 } 8
\tabulure 5 {4 1 5} {\whitetypes 4 {$x$} 2 1 5 \whitetypes 1 5 5 5 4 \whitetypes 5 4 2 2 2 } {9,13}
\tabulure 6 {2 2 2$_b$} {\whitetypes 2 1 5 5 1 \whitetypes 2 1 5 5 5 \blacktypes 2 5 5 {$x$} } {12}
\tabulure 7 {2$_b$ 2 1} {\blacktypes 2 5 5 1 \whitetypes 2 1 5 5 5 \whitetypes 1 5 5 5 {$x$} } {}
\tabulure 8 {2 1 5$_b$} {\whitetypes 2 1 5 5 5 \whitetypes 1 5 5 5 5 \blacktypes 5 5 4 2 } {11}
\tabulure 9 {5 4$_b$ 2} {\whitetypes 5 {$x$} 2 2 2 \blacktypes 4 2 1 1 \whitetypes 2 1 5 5 {$y$} }
                        {10,14}
\tabulure {10} {2$_b$ 1 1$_b$} {\blacktypes 2 5 5 5 \whitetypes 1 5 5 5 5 \blacktypes 1 5 5 {$x$} }
                        {}
\tabulure {11} {5$_b$ 4 2$_b$} {\blacktypes 5 {$x$} 2 2 \whitetypes 4 2 2 1 5 
                                \blacktypes 2 5 5 {$y$} } 7
\tabulure {12} {5 1$_b$ 5} {\whitetypes 5 {$x$} 2 2 2 \blacktypes 1 2 2 3 \whitetypes 5 3 2 2 2 }
                        {15}
\tabulure {13} {5 4$_b$ 2} {\whitetypes 5 {$x$} 2 2 2 \blacktypes 4 2 1 1 \whitetypes 2 1 5 5 {$y$} }
                        {10,14}
\tabulure {14} {1 1$_b$ 5} {\whitetypes 1 5 5 5 5 \blacktypes 1 5 5 4 \whitetypes 5 4 2 2 2 }
                        {13}
\tabulure {15} {2 3$_b$ 2} {\whitetypes 2 1 5 5 1 \blacktypes 3 1 5 1 \whitetypes 2 1 5 5 {$x$} }
                        {12}
}
\hfill}
}
\vskip 10pt
\cqfd



\end{document}